# Title: Stochasticity in Ca$^{2+}$ increase in spines enables robust and sensitive information coding

**Authors:** Takuya Koumura[1†], Hidetoshi Urakubo[2††], Kaoru Ohashi[2], Masashi Fujii[2] and Shinya Kuroda[1,2]*

**Affiliations:** [1]Undergraduate Department of Bioinformatics and Systems Biology, [2]Department of Biophysics and Biochemistry, Graduate School of Science, University of Tokyo, Hongo 7-3-1, Bunkyo-ku, Tokyo 113-0033, Japan.

†Present address: Department of Life Sciences, Graduate School of Arts and Sciences, University of Tokyo, Komaba 3-8-1, Meguro-ku, Tokyo 153-8902, Japan.

††Present address: Department of Systems Science, Graduate School of Informatics, Kyoto University, Uji 611-0011  JAPAN

**Running head:** Probabilistic coding of Ca$^{2+}$ increase in spines

**\* E-mail:** skuroda@bi.s.u-tokyo.ac.jp

## ABSTRACT

A dendritic spine is a very small structure (~0.1 $\mu m^3$) of a neuron that processes input timing information. Why are spines so small? Here, we provide functional reasons; the size of spines is optimal for information coding. Spines code input timing information by the probability of $Ca^{2+}$ increases, which makes robust and sensitive information coding possible. We created a stochastic simulation model of input timing-dependent $Ca^{2+}$ increases in a cerebellar Purkinje cell's spine. Spines used probability coding of $Ca^{2+}$ increases rather than amplitude coding for input timing detection via stochastic facilitation by utilizing the small number of molecules in a spine volume, where information per volume appeared optimal. Probability coding of $Ca^{2+}$ increases in a spine volume was more robust against input fluctuation and more sensitive to input numbers than amplitude coding of $Ca^{2+}$ increases in a cell volume. Thus, stochasticity is a strategy by which neurons robustly and sensitively code information.

**AUTHOR SUMMARY (100-200 words, 166 words)**

A spine is a small platform ($0.1~\mu m^3$, $10^4$-fold smaller than a cell) for information processing in a neuron. Biochemical reactions in a spine inevitably becomes noisy due to small numbers of molecules in a small volume. Such noise of reactions generally perturbs signal transmission, but how can a spine accurately receive input information under such noisy conditions? Here we show that a spine utilizes noise-enhancing signal transmission termed as 'probability coding' resulting from the smallness of a spine, and the spine volume appeared 'optimal' for the information coding with the fixed amount of materials and space. In the stochastic simulation of $Ca^{2+}$ increases in various volumes, the spine volume preferred the 'probability coding', whereas the cell volume preferred the general 'amplitude coding'. We found that information per volume was largest in a spine volume, and that the 'probability coding' was much more robust against noise than 'amplitude coding'. These findings clearly demonstrate the advantage of smallness of spines in information processing, especially for signals of synaptic plasticity.

## INTRODUCTION

A dendritic spine is a small structure on the dendrites of a neuron that processes input timing information from other neurons, and typically, tens to hundreds of thousands of spines are present on a neuron [1,2]. Small volumes and large numbers are representative characteristics of spines. For example, the volume of the spines at the parallel fibre (PF)-cerebellar Purkinje cell synapse is approximately 0.1 $\mu m^3$ [3], which is $10^4$-fold smaller than the cell body (5,000 $\mu m^3$) [4,5], and 175,000 spines are present on a single neuron (**Fig. 1A**) [1]. Cerebellar Purkinje cells receive two inputs: PF inputs from granular neurons that are thought to code sensorimotor signals and a climbing fibre (CF) input from inferior olivary nucleus that is thought to code error signal [6-8]. The timing between the PF-CF activation has been shown to be important for associative eyelid conditioning in cerebellar learning [9,10]. Coincident PF and CF inputs but not PF or CF inputs alone to spines at the PF-Purkinje cell synapse induce large $Ca^{2+}$ increases via IP$_3$ (inositol trisphosphate)-induced $Ca^{2+}$ release (IICR) from intracellular $Ca^{2+}$ stores, such as the endoplasmic reticulum (ER) [11,12] (**Fig. 1B**). Large $Ca^{2+}$ increases subsequently trigger long-term decreases of synaptic strength that are known as cerebellar long-term depression (LTD) [13], which is thought to be the molecular and cellular basis of our motor learning [7].

$Ca^{2+}$ increases depend on the relative timing of the PF and CF inputs, and the time window between PF and CF inputs that produces $Ca^{2+}$ increases has been experimentally determined (**Fig. 1C**) [12]. Large $Ca^{2+}$ increases occur only when PF and CF inputs are coincident at a given synapse within a 200 msec time window. We previously developed a deterministic kinetic model of PF and CF inputs-dependent $Ca^{2+}$ increases based on experimental findings and reproduced the time window curve of the $Ca^{2+}$ increase [14]. However, in a spine, the number of signaling molecules is limited to tens to hundreds (**Fig. 1B**); thus, the number of molecules should fluctuate due to stochastically occurring reactions. Indeed, experimentally, the same coincident PF and CF inputs do not always induce similar large $Ca^{2+}$ increases; in some cases, large $Ca^{2+}$ increases are observed, whereas in other cases, they are not (**Fig. 1C**) [12]. How can a spine robustly and sensitively code input timing information under such stochastic conditions with limited numbers of molecules [15-17]? To address this issue, we created a stochastic simulation model of input-timing-dependent $Ca^{2+}$ increase in a spine volume based on the deterministic kinetic model of $Ca^{2+}$ increase in a spine at the PF-Purkinje cell synapse [14] that incorporated stochastic reactions due to the small number of molecules.

## RESULTS

### Stochastic simulation of $Ca^{2+}$ increase in a spine at the PF-Purkinje cell synapse

We used the same kinetic framework and parameters of PF and CF inputs-dependent $Ca^{2+}$ increase in the stochastic simulation used in our previous deterministic model [14]: the PF inputs presynaptically released glutamate (**Fig. 1B**). Glutamate binds and activates metabotropic glutamate receptors (mGluRs), which leads to an increase in $IP_3$. Glutamate also binds and activates ±-amino-3-hydroxy-5-methyl-4-isoxazole propionic acid receptors (AMPARs), which leads to an increase in $Ca^{2+}$ influx through indirect activation of voltage-gated $Ca^{2+}$ channels (VGCC). The CF input causes $Ca^{2+}$ influx through VGCCs. The $IP_3$ receptor, an intracellular $Ca^{2+}$ channel, is activated only when it binds simultaneously with both $IP_3$ and $Ca^{2+}$. When the ternary complex of $Ca^{2+} \cdot IP_3 \cdot IP_3$ receptor is formed by the PF input, the $Ca^{2+}$ influx via CF input triggers a positive feedback loop of $Ca^{2+}$-dependent $Ca^{2+}$ release that leads to a large $Ca^{2+}$ increase from intracellular $Ca^{2+}$ stores, such as the ER. This large $Ca^{2+}$ increase induces a delayed negative feedback inhibition of $Ca^{2+}$ release from the stores at higher levels of $Ca^{2+}$.

Therefore, PF and CF inputs-dependent $Ca^{2+}$ increase is generated by positive- and delayed negative-feedback loops. This forms a single, large and transient $Ca^{2+}$ increase, depending on the relative timing of the PF and CF inputs.

We performed stochastic simulation of the $Ca^{2+}$ increase due to PF and CF inputs in a spine and in a cell volume (**Fig. 2, Fig. S1**). The volumes of spines and cells have been reported to be 0.1 $\mu m^3$ [3] and 5,000 $\mu m^3$ [4,5], respectively. We hereafter denote 0.1 $\mu m^3$ as a spine volume and 5,000 $\mu m^3$ as a cell volume. In response to coincident PF and CF inputs (" t = 160 msec), a large $Ca^{2+}$ increase was always observed in a cell volume (**Fig. 2A**, black lines). In a spine volume, a large $Ca^{2+}$ increase was observed in many (1371 cases in 2000 trials) but not all cases (**Fig. 2A**, gray lines). In response to PF and CF inputs with reversed timing (" t = −400 msec), no large $Ca^{2+}$ increase was observed in a cell volume (**Fig. 2B**, black lines). In a spine volume, however, a large $Ca^{2+}$ increase was observed in some cases (690 cases in 2000 trials), but not in many cases (**Fig. 2B**, gray lines). In both timing, the amplitudes of the $Ca^{2+}$ increases fluctuated between trials in a spine volume, but not in a cell volume. Regardless of the input timing, $Ca^{2+}$ increase occurred only once. In order to quantify the large $Ca^{2+}$ increase, we measured the response of $Ca^{2+}$ increase by logarithm of temporally integrated $Ca^{2+}$ subtracted by the basal $Ca^{2+}$ concentration (Materials and Methods). The probability distribution of the $Ca^{2+}$ response in a spine volume appeared bimodal (**Fig. 2C**). The narrow mode on the left corresponds with trials of no $Ca^{2+}$ increase, and the wide mode on the right corresponds with the trials with large $Ca^{2+}$ increases. In response to coincident PF and CF inputs, the height of the right mode increased and that of the left mode decreased. The mean of each mode appeared similar, regardless of the timing between the PF and CF inputs. This finding suggests that in a spine volume, the timing information is encoded by the probability of a large $Ca^{2+}$ increase (fraction of trials) rather than the amplitude of $Ca^{2+}$ increase.

**Probability coding of $Ca^{2+}$ increase in a spine**

We next examined the distribution of the $Ca^{2+}$ response in every 20 msec PF-CF intervals between −400 msec and 600 msec in a spine volume (**Fig. 3A**, left panel). In all PF-CF intervals, the probability distribution of the $Ca^{2+}$ response appeared bimodal. In the time window between 0 msec and +340 msec PF-CF intervals, the upper mode of the distribution appeared dominant. Outside of this time window, the lower mode of the distribution appeared dominant. Such bimodality was not observed at all in the distribution in a cell volume, where unimodality was always observed regardless of the PF-CF interval (**Fig. 3B**, left panel). In both a spine and a cell volume, however, the mean $Ca^{2+}$ response showed similar graded bell-shaped curves (**Fig. 1C**). These results suggest that information of the PF-CF interval is coded by the probability of a large $Ca^{2+}$ increase in a spine volume, and by the amplitude of a large $Ca^{2+}$ increase in a cell volume.

Therefore, we examined whether the probability or amplitude of the $Ca^{2+}$ increase codes the information on PF-CF interval by decomposing the distribution of the $Ca^{2+}$ response into the probability and amplitude components (**Fig. 2D**, Materials and Methods). First, we determined the threshold between the two peaks of the marginal distributions (**Fig. 3**, arrowheads in left panels). Then, we divided the distribution by the threshold (i.e. distributions above or below the threshold). The probability component of the distribution was defined as the frequencies of the $Ca^{2+}$ response in each divided distribution (**Fig. 2D**, middle panel). The amplitude component was defined as the distribution of the $Ca^{2+}$ response in each divided distribution (**Fig. 2D**, bottom panel). The total distribution of the $Ca^{2+}$ response (**Fig. 2D,** top panel) is the sum of the amplitude components weighted with the probability components (see Materials and Methods).

In a spine volume, the probability component of the $Ca^{2+}$ response was graded bell-shaped curves along the PF-CF interval (**Fig. 3A**, middle panel), whereas the amplitude component of the $Ca^{2+}$ response was similar regardless of the PF-CF interval (**Fig. 3A**, right panel). This result indicates that in the stochastic model in a spine volume, timing information between PF and CF inputs is coded by the probability component of the $Ca^{2+}$ response rather than the amplitude component of the $Ca^{2+}$ response.

In contrast, in a cell volume, the amplitude component of the $Ca^{2+}$ response showed a graded bell-shaped curve along the PF-CF interval (**Fig. 3B**, right panel). However, the probability component of the $Ca^{2+}$ response was a rectangular curve along the PF-CF interval (**Fig. 3B**, middle panel), meaning that the probability component can be used only to detect the binary input timing information on whether the input timing was inside or outside the time window; it cannot be used for the detection of the small input timing difference. In an intermediate volume, both the probability and amplitude components showed intermediate property between those of the spine and cell volume (**Fig. 3C**, middle and right panels). These results suggest that the probability and amplitude components of the $Ca^{2+}$ response are the dominant factor in the distribution of the $Ca^{2+}$ response in a spine volume and in a cell volume, respectively.

We next quantified the input timing information coded by the distribution of the $Ca^{2+}$ response, the probability components, and the amplitude components in a volume-dependent manner (**Fig. 3D**, **Fig. S2**). We first measured the total input timing information coded by the distribution of the $Ca^{2+}$ response by mutual information between the $Ca^{2+}$ response and PF-CF interval (**Fig. 3D**, black line). Then, we calculated the distribution without the probability component or amplitude component from the original distribution of the $Ca^{2+}$ response (**Fig. S2**, **Materials and Methods**). We measured the information coded by the probability component by the Kullback–Leibler (KL) divergence of the probability component-removed distribution from the original distribution of the $Ca^{2+}$ response (**Fig. 3D**, red line). Similarly, we measured the information coded by the amplitude component by the KL divergence of the amplitude component-removed distribution from the original distribution of the $Ca^{2+}$ response (**Fig. 3D**, blue line). The input timing information coded by the probability component, amplitude component, and original distribution increased monotonically as the volume became larger (**Fig. 3D**). As the volume increased, the information coded by the probability component gradually increased and converged to 1bit in a cell volume, indicating that a cell can discriminate binary information with the probability component (**Fig. 3D**, red line). The information coded by the amplitude component was lower than that coded by the probability component in a spine volume, and became larger than that coded by the probability component in the volume larger than 10 $\mu m^3$ (**Fig. 3D**, blue line). The total information on input timing in a cell volume was larger than that in a spine volume; however, a spine is much smaller than a cell in the first place. To fairly compare the information coding capability across volumes, we divided the mutual information by the volume and obtained the mutual information per volume (**Fig. 3E**). The mutual information per volume was the largest in a volume of 0.5 $\mu m^3$, which suggests that a spine has the nearly optimal volume for the information coding in terms of the input timing information per volume. In other words, input timing information is coded most efficiently in the volume around a spine volume if given the fixed amount of materials and space.

Next, we examined the relative contribution of the probability or amplitude component to the total information (**Fig. 3F**). We divided the information coded by probability or amplitude component by the sum of them, which is equal to the total information. The contribution of the probability component was larger than that of amplitude component in the volume smaller than 8 $\mu m^3$. This indicates that the probability coding was dominant in a spine. Together with the result of efficient information coding per volume in a spine, the probability coding in this study could be one of stochastic facilitation, the phenomenon in which information transmission is enhanced by stochasticity [18]. In contrast, in the larger volume, the relative contribution of the amplitude component became larger than that of the probability component, indicating that amplitude coding is dominant. Thus, a spine codes input timing information by the probability component of the distribution of the $Ca^{2+}$ response by utilising its stochasticity which derives from its smallness, and a cell codes input timing information by the amplitude component of the distribution. In addition, this suggests that the same IICR system, a $Ca^{2+}$ increase generator, can switch between probability coding and amplitude coding for input timing detection depending on the volume. As volume in the stochastic simulation increased, the behaviour of the stochastic model became similar to that of the deterministic model, which has a volume that can be regarded as infinity. Indeed, the stochastic simulation in a cell volume (5,000 $\mu m^3$) showed almost the same response as the previous deterministic model (**Fig. 2A, B, Fig. S1**) [14]. Therefore, we hereafter used the deterministic model as the model in a cell volume.

**Robust and sensitive information coding in a spine**

Since glutamate release from PFs fluctuates between trials in the physiological condition [19], the $Ca^{2+}$ increase should be affected by the fluctuation of the amplitude of PF inputs (PF amplitude). We simulated the $Ca^{2+}$ increase with the fluctuation of PF amplitude between trials, and asked whether the probability coding in spines are advantageous for robust information coding on input timing against fluctuation of PF amplitude (**Fig. 4A, B**). Here we use the term "robustness" of information coding as invariability of input timing information against the increase of the fluctuation of PF amplitude. The distributions of $Ca^{2+}$ response in a spine volume were similar regardless of the coefficient of variation (CV) of the PF amplitude (**Fig. 4A**, upper panels). However, the distribution of $Ca^{2+}$ response in a cell volume appeared to be affected by the CV of the PF amplitude (**Fig. 4A**, lower panels). In a cell volume, distribution of $Ca^{2+}$ response showed larger variance if CV of PF amplitude were increased from 0.1 (**Fig. 4A**, lower left panel) to 0.5 (**Fig. 4A**, lower right panel). These results suggest that input timing information coding is more robust in a spine volume than in a cell volume. Indeed, the input timing information coded by the $Ca^{2+}$ response in a spine volume remained almost constant regardless of the CV of the PF amplitude (**Fig. 4B**, red line). The input timing information decreased only by 14.3% when the CV of the PF amplitude increased from 0.05 to 0.5 (**Table 1**). In contrast, the input timing information in a cell volume largely decreased as the CV of the PF amplitude increased (**Fig. 4B**, black line); it decreased by 76.9% when the CV of the PF amplitude increased from 0.05 to 0.5 (**Table 1**). These results indicate that spines utilize their smallness for the robust input timing information coding. The fluctuation in the amplitude of EPSPs has been reported to be 50% [19], which corresponds with a CV of 0.5. This fluctuation may be overestimated due to the inclusion of postsynaptic fluctuation; therefore, we used a CV of 0.1 for further study.

We also examined the sensitivity of the input timing information to the number of PF inputs in spines and in a cell (**Fig. 4C, D**). Here we use the term "sensitivity" of information coding as invariability of input timing information against the decrease of the number of PF inputs. In response to a single PF input, large $Ca^{2+}$ increase was not observed both in a spine and cell volume (**Fig. 4C**, left panels). In response to three PF inputs, large $Ca^{2+}$ increase appeared in a spine volume with PF-CF intervals around 200 ms, whereas large $Ca^{2+}$ increase was not observed at all in a cell volume (**Fig. 4C**, middle panels). In response to seven PF inputs, large

$Ca^{2+}$ increase was observed both in a spine and cell volume (**Fig. 4C**, right panels). These results

suggest that input timing information coding is more sensitive to the number of PF inputs in a

spine volume than in a cell volume. Indeed, when the number of PF inputs was decreased from 5

to 3, input timing information decreased only by 36.1% in a single spine, whereas it decreased by

40.6% in a cell (**Table 1**). Taken together, information coding in a spine volume was more

robust against the fluctuation of the PF amplitude and more sensitive to the small number of the

PF inputs than that in a cell volume.

**Testable Prediction**

Our results imply that the amplitude of a PF input, which corresponds to concentration of a glutamate pulse, is coded by the probability of large $Ca^{2+}$ increase in a spine volume and the amplitude of $Ca^{2+}$ in a cell volume. This indicates that the concentration of a glutamate pulse should be observed by the number of spines with large $Ca^{2+}$ increases in spines, whereas they should be observed by large $Ca^{2+}$ response in a soma whose volume corresponds to a cell volume.

Based on this, we propose the following testable prediction (**Fig. S4**). Based on the simulation results, we provided one of the predicted examples. With repetitive application of a glutamate pulse with the same concentration, a similar level of $Ca^{2+}$ increase is always expected in a soma, whereas a similar number of spines with large $Ca^{2+}$ increase is expected in spines (**Fig. S4A**).

Importantly, the spines with large $Ca^{2+}$ increases and the amplitudes of $Ca^{2+}$ increases in each

spine are different between trials. This result will support the idea that generation of large $Ca^{2+}$

increases in spines is stochastic rather than deterministic. Upon repetitive application of glutamate pulses of various concentrations, the number of spines with large $Ca^{2+}$ increase and the amplitude of $Ca^{2+}$ response in soma are predicted to increase as the concentration of glutamate increases (**Fig. S4B**). The spines with large $Ca^{2+}$ increases are different between trials, even at the same concentration of glutamate. Thus, repetitive application of glutamate pulses at various doses to a Purkinje cell will reveal the distinct characteristics of $Ca^{2+}$ increases in spines and the soma. Importantly, at lower doses of glutamate, large $Ca^{2+}$ increases were observed in spines but not in the soma, indicating that spines are more sensitive to glutamate than the soma (**Fig. S4C**). This will support our finding of increased sensitivity of spines to input (**Fig. 4C, D**).

**DISCUSSION**

One of the features of our stochastic model is that the reactions and parameters of the deterministic model, which were estimated from experimental data [14], were directly used. Importantly, we created the stochastic model without changing any reactions or parameters of the deterministic model and obtained the result that the volume of real spines in Purkinje cells is nearly optimal for information coding. Although the information per volume in a spine volume was not the largest, it was almost the same as the largest value in a volume of 0.5 µm$^3$ and far larger than that in a cell volume. Thus, our results suggest that spines can code input timing information most efficiently if given the fixed amount of materials and space. This supports the idea that the small volume of spines has been evolutionally selected based on biochemical reactions and kinetic parameters.

We provided a functional reason for the small size of spines. The smallness of spines was optimal for maximising input timing information per volume, and spines were more robust and sensitive to input than a cell. In this study, the source of fluctuation was derived from stochasticity in the number of molecules. As the number of molecules in a spine becomes smaller, the fluctuation in the number becomes larger. For example, the CVs of the $Ca^{2+}$ indicator in a cell volume and in a spine volume are 0.003 and 0.72 at 160 msec PF-CF interval, respectively. Indeed, it has been reported that CV of the peak amplitude of dendritic $Ca^{2+}$ transient in cerebellar Purkinje cells *in vivo* was 0.4 [20]. The fluctuation of the number of molecules is very large in a spine, suggesting that the number of molecules (amplitude) is not suitable as an information carrier. Such small numbers of molecules in a spine suggest that there may be many defective spines that do not contain essential molecules, and the reliability of a single spine is very low compared with that of a cell that contains huge numbers of molecules. Despite the low reliability of a single spine, summation of the probability of large $Ca^{2+}$ increases in many spines may overcome the amplitude of $Ca^{2+}$ in a cell in terms of the reliability of information coding. Thus, the smallness and numerosity of spines may be a unique strategy for robust and sensitive information coding in neurons. Spines in other types of neurons, and small and numerous organelles may employ a similar strategy using probability coding. Our finding also raises the caution that in contrast to experiments at the scale of the cell, in experiments at the scale of a spine, the probability of signaling activities over many trials or in many spines should be measured and interpreted rather than measuring only the amplitude in a single trial in a single spine. Note that the term 'probability coding' has the different meaning from that used in the field of neural dynamics of decision-making, in which it is used as the coding of the probability of expected event such as reward by neural dynamics [21,22].

In this study, the input timing information was encoded into the probability of $Ca^{2+}$ increases. The next question is how timing-information across the spines is decoded to the final output, cerebellar LTD? Experimentally, cerebellar LTD is induced by conjunctive activation of a PF bundle and CFs, and the responses from many PF synapses are simultaneously observed [13]. As it has recently been reported that all-or-none LTD at many synapses can show a graded response in a stochastic situation [19,20], it is likely that cerebellar LTD show a graded bell-shaped time-window curve. Similar phenomenon has been observed in synaptic plasticity at hippocampal synapses experimentally [23-25].

Furthermore, the input timing information can be decoded even in LTD at a single synapse Induction of cerebellar LTD has been reported to be regulated by a positive feedback loop between PKC and MAP kinase [26] and modelled all-or-none deterministic models [27]; however, it has recently been reported that cerebellar LTD in a single spine can show a graded response rather than an all-or-none response in a stochastic situation [28,29]. It is likely that cerebellar LTD show a graded bell-shaped time-window curve, indicating that cerebellar LTD can decode input timing information from the frequency of $Ca^{2+}$ increases.

The large $Ca^{2+}$ increases are generated by the IICR system in cerebellar Purkinje cells [30-32]. The IICR system is an excitable system with a positive feedback loop (in which gain is controlled by $IP_3$) and delayed negative feedback (**Fig. 1B**) [14]. The IICR system switches information-coding modes depending on volume; probability coding is used at spine volumes, and amplitude coding is used at cell volumes. We examined a mechanistic insight of the IICR system that generates stochastic facilitation (**Fig. S3**). Blocking the interaction of $IP_3$ with the $IP_3$ receptor, which controls the feedback gain of the regenerative cycle of $Ca^{2+}$, resulted in disappearance of the large $Ca^{2+}$ increases above the threshold and decrease of the amplitude of $Ca^{2+}$ below the threshold (**Fig. S3B**). Blocking the interaction of $Ca^{2+}$ with the $IP_3$ receptor, which triggers the regenerative cycle of $Ca^{2+}$, resulted in disappearance of the large $Ca^{2+}$ increases above the threshold (**Fig. S3C**). Blocking $Ca^{2+}$ influx through the $IP_3$ receptor, which is responsible for regenerated increase of $Ca^{2+}$, resulted in the disappearance of the large $Ca^{2+}$ increases above the threshold and decrease in the amplitude of $Ca^{2+}$ below the threshold (**Fig. S3D**). Thus, the regenerative positive feedback loop in which gain is controlled by $IP_3$ appeared to be an essential mechanism for probability coding.

The IICR system is an excitable system, but its mechanism is different from that of conventional excitable systems, such as the relaxation oscillator [33]; the relaxation oscillator has a fixed threshold and stable point, whereas the IICR system has a controllable threshold and stable point. This may make a difference in a stochastic situation; however, this possibility remains to be elucidated. This makes a difference in a deterministic situation. The relaxation oscillator shows an all-or-none response with a fixed amplitude, whereas the IICR system shows a graded response with a controllable amplitude. The relaxation oscillator can transform input strength into oscillation frequency [33] rather than amplitude, whereas the IICR system can transform input strength into amplitude of $Ca^{2+}$ increase in a cell volume.

It has recently been shown that the IICR system may use probability coding of $Ca^{2+}$ increases even in a cell volume [34-37]. Such probabilistic $Ca^{2+}$ increase may arise from stochastic individual $Ca^{2+}$ release through $Ca^{2+}$ channels in microdomains [35-37]. Even in a cell volume, probabilistic coding due to small numbers of molecules can be realized. This may contribute to efficient information coding in a cell volume.

Noise-dependent enhancement of the detection of weak information-carrying signals in a threshold system is known as stochastic resonance. Stochastic resonance in a narrow sense, or classical stochastic resonance, is a phenomenon in which detection of weak periodic signals by nonlinear threshold systems is enhanced by exogenous additive white noise [18]. IICR system is similar to the classical stochastic resonance; however, it has different characteristics; the source of noise is different. In classical stochastic resonance, noise is exogenous and not correlated with the signal amplitude. In contrast, in the IICR system, noise is produced by the intrinsic stochasticity of chemical reactions and correlated with the number of molecules. Thus, efficient information coding in IICR system can be referred as stochastic resonance in a broad sense. As a generalized concept of the classical stochastic resonance, or stochastic resonance in a broad sense, the term "stochastic facilitation" has been recently proposed [18]. Stochastic facilitation is a concept in which biologically relevant noise enhances the efficiency and/or effectiveness of the system. Therefore efficient information coding in the IICR system is one of the example of stochastic facilitation.

**Materials and Methods**

**Stochastic model**

We created a stochastic simulation model of input timing-dependent $Ca^{2+}$ increases based on the deterministic kinetic model of $Ca^{2+}$ increases in cerebellar Purkinje cells [14]. We used the following molecules, reactions and parameters for the stochastic model (which were identical to those used for our previous deterministic model): 56 molecules, 43 reactions and 96 parameters. Among the 96 parameters, 60 parameters were determined by experiments in the deterministic kinetic model [14]. The rest of the parameters were estimated by fitting the model to the experimental data that contain information concerning the parameters [12]. Thus, the parameters in the model are likely to be biologically plausible. All enzymatic reactions based on the Mechaelis-Menten formulation was decomposed into three elementary one-way reactions with the parameters $k_1$, $k_{-1}$, and $k_{cat}$, and $k_1$ was assumed to be four times larger than $k_{cat}$ [14]. This model has two types of inputs: a CF input and 5 PF inputs, which lead to activation of AMPARs and VGCCs. Instead of modeling the kinetics of AMPARs and VGCCs, a rectangular $Ca^{2+}$ pulse was applied to the cytosol to achieve a $Ca^{2+}$ influx as a result of VGCC opening, because the period of VGCC opening is known to be very short, less than ''t = 20 msec [38]. The CF input give a pulse of $Ca^{2+}$ influx (83.3 µM/msec for 2 msec) as $Ca^{2+}$ influx through VGCCs. The 5 PF inputs give 5 pulses of glutamate influx (5 µM/msec for 1 msec for each pulse) and 5 pulses of $Ca^{2+}$ influx (25 µM/msec for 1 msec for each pulse) at 100 Hz, the latter of which were regarded as $Ca^{2+}$ influx through VGCCs activated by AMPARs.

**Numerical simulation**

In the deterministic model, each reaction was represented as an ordinary differential equation (ODE) as was conducted in the previous model [14]. The ODEs were numerically solved by the Bogacki–Shampine method [39], a variation of Runge-Kutta method, with adaptive time step control ranging from 0.1 to 10 μsec. In the stochastic model, we discretized the number of molecules in the ODE model. The initial numbers of molecules in the stochastic model were determined as integers based on the initial concentrations in the deterministic model.

In general, surface area of the membrane is proportional to the order of the square of length, whereas cell volume is proportional to the cube of that, which means that, as a system size increases, the increase rate of number of membrane molecules becomes smaller than that in cytoplasm. Hence, in the cases of larger systems than the spine, the number of membrane molecules that can activate the cytosolic molecules is so small that most of substrates are not activated by the stimulation. Actually, when a volume was 8- or 125-fold larger than a spine volume, large $Ca^{2+}$ increase did not occur any more at any PF-CF intervals (**Fig. S7**). In order to uncover the simple influences of the smallness of a spine and the number of molecules, it is required that the effect of the stimulation for a cell on the cytosolic molecules through the membrane protein is transferred as well as that for a spine. Therefore, we assumed that the number of membrane proteins is proportional to the volume, *i.e.* the cube of length, throughout this study.

Another reason for this assumption is that, if the number of membrane proteins is to be set proportional to the surface area, the unknown parameters whose values were determined by fitting the simulation outputs to the experimental results in the previous deterministic model [14] must be re-determined in each volume. This makes the comparison of results across the volume unfair. Therefore, since in this study we wanted to focus only on the effects of the volume itself, we set the number of all molecules proportional to the volume.

To solve the stochastic model numerically, we utilized the modified tau-leaping method, which is an approximation of the stochastic simulation algorithm (SSA) [40]. Despite the simplicity of the method by Cao *et al.* [37], their method provides good approximations in the cases of the low-order-reaction-systems like this study [38]. We validated the use of modified tau-leaping method by comparing the results derived from modified tau-leaping method (**Fig. 2A-C**) with that derived from SSA (**Fig. S5A-C**). We repeated the same simulation with different random seeds. The number of simulation trials were as follows: 8,000 trials in the volume of 0.05 μm$^3$, 2,000 trials in the volume of 0.1 μm$^3$, 1,000 trials in the 0.5 μm$^3$, 0.8 μm$^3$, and 1 μm$^3$ volumes, 100 trials in the 10 μm$^3$, 12.5 μm$^3$, and 100 μm$^3$ volumes, and 20 trials in the 1,000 μm$^3$ and 5,000 μm$^3$ volumes. The simulation for the testable prediction was conducted 100 times. The numbers of trials were confirmed to be sufficient by resampling the simulated data (**Fig. S5D**). The simulation result in a cell volume was nearly identical to that in the previous deterministic model, supporting the validity of the stochastic model.

**Decomposition of probability and amplitude components from the distribution of Ca$^{2+}$ response**

We measured the response of Ca$^{2+}$ increase, *Ca$_{res}$*, by logarithm of temporally integrated Ca$^{2+}$ subtracted by the basal Ca$^{2+}$ concentration:

$$Ca_{res} = \log_{10}[\frac{1}{2}\int_{-0.5}^{1.5}([\text{Ca}^{2+}](t) - [\text{Ca}^{2+}]_{\text{basal}})dt]\,, (1)$$

where [Ca$^{2+}$](t) and [Ca$^{2+}$]$_{\text{basal}}$ denote intracellular Ca$^{2+}$ concentration at time *t* and basal Ca$^{2+}$ concentration, respectively. We used logarithmic scale for Ca$^{2+}$ concentration assuming that downstream molecules of Ca$^{2+}$ respond linearly to the log-scaled stimulus. This assumption is justified by the fact that serially diluted solution is often used as stimulus in experiments. The probability distribution of $Ca_{res}$ over trials was expressed by $P(Ca_{res} | \Delta t)$ where $\Delta t$ denotes the PF-CF interval and the time difference between the first PF and CF inputs. The marginal distribution over PF-CF interval, $P(Ca_{res})$, was bimodal. Therefore, we set a threshold at the minimum between the two peaks of the marginal distribution (**Fig. 3A-C**, left panels). We decomposed the distribution of $Ca_{res}$ into the amplitude and probability components:

$$P(Ca_{res} | \Delta t) = P(s = 0 | \Delta t)P(Ca_{res} | s = 0, \Delta t) + P(s = 1 | \Delta t)P(Ca_{res} | s = 1, \Delta t) , (2)$$

where $s = 1$ represents that $Ca_{res}$ exceeded the threshold, and $s = 0$ otherwise (**Fig. 2D**). In this representation, $P(s | \Delta t)$ and $P(Ca_{res} | s, \Delta t)$ corresponds to the probability and amplitude component, respectively. The probability component, $P(s | \Delta t)$, gives the probabilities of the $Ca_{res}$ above/below the threshold, and the amplitude component, $P(Ca_{res} | s, \Delta t)$, gives the probability distributions of $Ca_{res}$, partitioned by the threshold.

**Calculation of input timing information**

We measured the input timing information coded by the $Ca^{2+}$ response as mutual information (MI) between the $Ca_{res}$ and PF-CF interval:

$$I_{total} = \sum_{\Delta t} P(\Delta t) \int_{Ca_{res}} P(Ca_{res} | \Delta t) \log_2 \frac{P(Ca_{res} | \Delta t)}{P(Ca_{res})} dCa_{res} . (3)$$

Here, $P(\Delta t)$ follows the uniform distribution, and $\Delta t$ denotes the PF-CF interval ($\Delta t = -400, -380, -360, \ldots, +600$ msec). We used the uniform distribution for $P(\Delta t)$ because, in physiological or behavioral experiments, $P(\Delta t)$ totally depends on the experimental conditions, and $P(\Delta t)$ largely differ depending on task, situation, and many other factors. Therefore, optimal $P(\Delta t)$ cannot be uniquely determined. We computed the mutual information with $P(\Delta t)$ of Gaussian distribution, and confirmed that changing $P(\Delta t)$ does not affect the results qualitatively (**Fig.**

**S8**). In the calculation of the MI, we divided the entire range of $Ca_{res}$ from minimum to maximum into 50 bins, where the intervals were determined based on the method described in Cheong *et al.* [ref], who estimated the MI taking the limit as the sample number approaches infinity. Note that for the probability density function, the MI (and also Kullback-Leibler (KL) divergence) is invariant under the any transformation of random variables. In this study, we estimated the MI by the abovementioned method, therefore, we could obtain almost the same value of MI for any transformation of $Ca_{res}$ (**Fig. S6**).

We also measured the information coded by the probability component by use of the KL divergence:

$$I_{prob} = \sum_{\Delta t} P(\Delta t) \int_{Ca_{res}} P(Ca_{res} \mid \Delta t) \log_2 \frac{P(Ca_{res} \mid \Delta t)}{P_{-prob}(Ca_{res} \mid \Delta t)} dCa_{res} \,, (4)$$

where $P_{-prob}(Ca_{res} \mid \Delta t)$ denotes the distribution of $Ca_{res}$ without the probability component, which was calculated by marginalizing $\Delta t$ out of the probability component $P(s \mid \Delta t)$ in $P(Ca_{res} \mid \Delta t)$ :

$$P_{-prob}(Ca_{res} \mid \Delta t) = P(s = 0)P(Ca_{res} \mid s = 0, \Delta t) + P(s = 1)P(Ca_{res} \mid s = 1, \Delta t) \,. (5)$$

Further, we measured the information coded by the amplitude component:

$$I_{amp} = \sum_{\Delta t} P(\Delta t) \int_{Ca_{res}} P(Ca_{res} \mid \Delta t) \log_2 \frac{P(Ca_{res} \mid \Delta t)}{P_{-amp}(Ca_{res} \mid \Delta t)} dCa_{res} \,, (6)$$

where $P_{-amp}(Ca_{res} \mid \Delta t)$ denotes the distribution of $Ca_{res}$ without the amplitude component, which was calculated by marginalizing $\Delta t$ out of the amplitude component $P(Ca_{res} \mid s, \Delta t)$ in $P(Ca_{res} \mid \Delta t)$ :

$$P_{-amp}(Ca_{res} \mid \Delta t) = P(s = 0 \mid \Delta t)P(Ca_{res} \mid s = 0) + P(s = 1 \mid \Delta t)P(Ca_{res} \mid s = 1) \,. (7)$$

Here, the sum of $I_{prob}$ and $I_{amp}$ is equal to $I_{total}$ because

$P(Ca_{res} < threshold \mid s = 1, \Delta t) = 0$ and $P(Ca_{res} > threshold \mid s = 0, \Delta t) = 0$ can be naturally

assumed.

The relative contributions of the probability and amplitude components to the input timing information were defined by $I_{prob}/I_{total}$ and $I_{amp}/I_{total}$, respectively.

**Fluctuation of PF input amplitude**

For fluctuation of PF input amplitudes, the concentration of 5 glutamate pulses and 5 $Ca^{2+}$ pulses fluctuated between trials under the fixed mean concentration (5 µM). The fluctuation of PF input amplitudes follows the zero-truncated normal distribution with a given coefficient of variation where the PF input amplitude must be over zero.

**Acknowledgements:** We thank Keiko Tanaka (KIST, Korea) for providing the image of cerebellar Purkinje cells and critically reading this manuscript. We also thank our laboratory members for critically reading this manuscript and their technical assistance with the simulation. The stochastic and deterministic models have been placed on our website (http://wwwkurodalab.org/info/Ca_Increases).

**Figure Legends**

**Figure 1. Ca$^{2+}$ increase in cerebellar Purkinje cells. A**, Cerebellar Purkinje cells. Male mouse cerebellar Purkinje cells were doubly stained with anti-calbindin antibody to visualize whole cells (green) and with anti-mGluR (metabotropic glutamate receptor) antibody to specifically visualize the spines of the PF-Purkinje cell synapses (red). The inset in the left image is magnified in the right panel. The average volume of spines in cerebellar Purkinje cells has been reported to be 0.1 µm$^3$, which is 10$^4$-fold smaller than a cell body (5,000 µm$^3$). White circles in the left and right panels indicate a typical soma and spine, respectively. **B**, Schematic representation of PF and CF inputs-dependent Ca$^{2+}$ increase in the simulation model[14] (Materials and Methods). Abbreviations: Glu; glutamate, mGluR; metabotropic glutamate receptor, IP$_3$; inositol trisphosphate; AMPAR; ±-amino-3-hydroxy-5-methyl-4-isoxazole propionic acid receptor, VGCC; voltage-gated Ca$^{2+}$ channels. Parentheses indicate initial numbers of the indicated molecules in the stochastic model. **C**, Ca$^{2+}$ responses along the PF-CF interval in the experiments (black circles)[12], and mean Ca$^{2+}$ responses in the stochastic simulation in a spine volume (0.1 µm$^3$) (solid line) and in the stochastic simulation in a cell volume (5,000 µm$^3$) (dashed line). The Ca$^{2+}$ response was defined as the average relative change in fluorescence (”F/F$_0$) of Magnesium Green 1, a Ca$^{2+}$ indicator[14]. Positive sign of the interval are given to ”$t$ msec when PF inputs precede CF inputs; otherwise, a negative sign is given. Five PF inputs at 100Hz and a single CF were given.

**Figure 2. Ca$^{2+}$ increases in the stochastic model. A, B,** Ca$^{2+}$ increase due to stimulation of PF and CF inputs with ”t = 160 msec (**A**) and ”t = −400 msec (**B**) in the stochastic model in a spine volume (gray lines, n=2,000 for each timing, 20 examples are shown) and in a cell volume (black

lines, n=20 for each timing). **C**, Distributions of the $Ca^{2+}$ response in a spine volume with ˝ t = 160 msec (solid line) and ˝ t = −400 msec (dashed line). Black and white arrowheads indicate the means of the $Ca^{2+}$ response in a cell volume with ˝ t = 160 msec and ˝ t = −400 msec, respectively. Note that $Ca^{2+}$ increases in a cell volume were almost the same between trials, leading to the overlap of the time courses (**A**, **B**). **D**, The distribution of the $Ca^{2+}$ response (upper panel) was divided into the distribution with $Ca^{2+}$ spikes ($s$=1, black) and the distribution without $Ca^{2+}$ spikes ($s$=0, gray with dashed line). Then, each right ($s$=1) and left ($s$=0) distribution is decomposed into the probability components, which is probability of $Ca^{2+}$ spiking (middle panel), and amplitude components, which is distribution of the $Ca^{2+}$ response conditioned by $Ca^{2+}$ spiking (lower panel). See materials and methods for the detailed descriptions.

**Figure 3. Volume-dependent time window of $Ca^{2+}$ increases. A-C,** Distributions of the $Ca^{2+}$ response along the PF-CF interval (left) in 0.1 µm$^3$ (a spine volume) (**A**), in 5,000 µm$^3$ (a cell volume) (**B**), and in 10 µm$^3$ (**C**). The probability components (middle) and the amplitude components (right) are also shown. The colors in the left and right panels code the probability of the $Ca^{2+}$ response at the indicated PF-CF interval. Arrowheads in the left panels indicate the thresholds between two peaks of the bimodal marginal distributions. The probability components denote the frequencies of the $Ca^{2+}$ response above/below the thresholds (solid/dashed lines, respectively), and the amplitude components denote the $Ca^{2+}$ response above/below the thresholds (indicated by solid/dashed braces, respectively). The PF-CF interval was defined by the time difference between the first PF and CF inputs. **D**, Volume-dependency of the input timing information coded by the total distribution of the $Ca^{2+}$ response (black), by the probability component (red), and by the amplitude component (blue). **E**, Volume-dependency of the input

timing information per volume. **F,** Relative contribution of the probability (red) and amplitude (blue) component to the input timing information.

**Figure 4. Robust and sensitive probability coding in the stochastic model. A,** Distribution of the $Ca^{2+}$ response in a single spine (upper panels) and in a cell (lower panels) due to stimulation of PF and CF inputs with fluctuating the PF amplitudes. CVs of the PF amplitudes were 0.1 (left panels) and 0.5 (right panels). **B**, Input timing information per volume, coded by the $Ca^{2+}$ response, in a spine (red) and in a cell (black). **C,** Distribution of the $Ca^{2+}$ response in a single spine (upper panels) and in a cell (lower panels) with a single PF input (left panels), three PF inputs (middle panels), seven PF inputs (right panels). In a spine, large $Ca^{2+}$ increase were observed in a few trials in response to three PF inputs (white arrowhead). CV of the PF amplitudes was 0.1. **D**, Input timing information per volume in a spine (red) and in a cell (black).

|  | Decrease of the input timing information | |
|---|---|---|
|  | CV of PF amplitude from 0.05 to 0.5 | # of PF inputs from 5 to 3 |
| spine | 14.3% | 36.1% |
| cell | 76.9% | 40.6% |

**Table 1. Decrease of the input timing information with the change of the CV of PF amplitude from 0.05 to 0.5 and the number of PF inputs from 5 to 3.**

**Supporting Information Legends**

**Figure. S1 | Time courses of concentrations of inositol trisphosphate (IP$_3$) and Ca$^{2+}$ in the endoplasmic reticulum (ER). A, B,** Time courses of concentrations of IP$_3$ in response to PF and CF inputs with ″t = 160 msec (**A**) and ″t = • 400 msec (**B**) in the stochastic model in a spine volume (gray lines, n=2,000 for each timing, 20 examples are shown) and in a cell volume (black lines, n=20 for each timing), respectively. **C, D,** Time courses of concentrations of Ca$^{2+}$ in the ER in response to PF and CF inputs with ″t = 160 msec (**C**) and ″t = • 400 msec (**D**) in the stochastic model in a spine volume (gray lines, n=2,000 for each timing, 20 examples are shown) and in a cell volume (black lines, n=20 for each timing), respectively.

**Figure. S2 | Distribution of Ca$^{2+}$ response and that without the probability and/or amplitude components. A,** Distribution of Ca$^{2+}$ response. **B, C,** Distribution of Ca$^{2+}$ response without probability (**B**) or amplitude (**C**) component. **D,** Distribution of Ca$^{2+}$ response without both components. Input timing information coded by the probability and amplitude component was calculated by the Kullback–Leibler (KL) divergence of the distribution of **B** and **C** from that of **A**, respectively. Input timing information coded by the distribution of Ca$^{2+}$ response was calculated by the mutual information between Ca$^{2+}$ response and PF-CF interval, which is equal to the KL divergence of the distribution of **D** from that of **A**, and also equal to the sum of the input timing information coded by probability and amplitude component.

**Figure. S3 | Mechanism of probability coding. A,** Schematic representation of the pathway deleted in the following figures. **B-D,** Distribution of $Ca^{2+}$ response along the PF-CF interval in a spine volume with blocking of the interaction of $IP_3$ with the $IP_3$ receptor (**B**), with blocking of the interaction of $Ca^{2+}$ with the $IP_3$ receptor (**C**), and with blocking of $Ca^{2+}$ influx through the $IP_3$ receptor (**D**) in the stochastic simulation. **E,** Distribution of $Ca^{2+}$ response without deletion (control).

**Figure. S4 | Possible experimental tests of probability coding in spines and amplitude coding in a soma. A,** $Ca^{2+}$ increase in response to repetitive addition of glutamate pulses with the same concentration in the stochastic model (spines) and in the deterministic model (soma). Colours code the concentration of $Ca^{2+}$. Darker colours indicate higher concentrations of $Ca^{2+}$. The time course of $Ca^{2+}$ increase in the deterministic model and stochastic model are shown. **B,** $Ca^{2+}$ increases in response to glutamate pulses of various concentrations in the stochastic model (spines) and the deterministic model (soma). Darker colours indicate higher concentrations of $Ca^{2+}$. The time course of $Ca^{2+}$ increase in the deterministic model and stochastic model are shown. **C,** Glutamate dose-responses curves of the fraction of spines with large $Ca^{2+}$ increase above the threshold in the stochastic model (solid lines) and $Ca^{2+}$ response in the deterministic model (dashed lines). Note glutamate inputs with amplitude CVs of 0.1 were used.

**Figure. S5 | Validation of the numerical simulation. A, B,** $Ca^{2+}$ increase due to stimulation of PF and CF inputs with ″ t = 160 msec (**A**) and ″ t = • 400 msec (**B**) by the stochastic simulation algorithm (SSA) (gray lines, n=2,000 for each timing, 20 examples are shown). **C,** Distributions

of the Ca$^{2+}$ response in a spine volume with ”t = 160 msec (solid lines) and ”t = • 400 msec

(dashed line) by the modified tau-leaping method (black lines, same as **Fig. 2C**) and SSA (red

lines). The correlation coefficients (CCs) of the red and black lines were calculated. The large

CCs demonstrates the validity of the modified tau-leaping method.

**Figure. S6 | Input timing informations calculated in normal scale.** Here we calculated the

input timing informations in normal scale of Ca$^{2+}$ response, and compared them with the input timing information calculated in logarithmic scale (**Fig. 3** and **4**). Their similarity demonstrates that the results were qualitatively not affected whether they were calculated in logarithmic scale or normal scale. **A,** Volume-dependency of the input timing information coded by the total distribution of the Ca$^{2+}$ response (black), by the probability component (red), and by the amplitude component (blue). **B,** Volume-dependency of the input timing information per volume.

**C,** Relative contribution of the probability (red) and amplitude (blue) component to the input timing information. **D,** Input timing information per volume, coded by the Ca$^{2+}$ response, in a spine (red) and in a cell (black). **E,** Input timing information per volume in a spine (red) and in a cell (black). **Fig. S6A**, **B**, and **C** correspond to **Fig. 3D**, **E**, and **F**, respectively. **Fig. S6D** and **E** correspond to **Fig. 4B** and **D**, respectively.

**Figure. S7 | Results of simulation under the condition that the numbers of membrane**

**molecules were proportional to the surface area.** In this study, we assumed that the numbers of membrane molecules were proportional to the volume of the system. Here, to check the results of the simulation with the numbers of membrane molecules proportional to the surface area, we changed the model and performed the stochastic simulation. In this model, we altered the following three types of parameters: (i) initial numbers of the membrane molecules, (ii) membrane permeability coefficients, and (iii) constants for the propensity functions of the bimolecular and trimolecular reactions occurring on the membrane. (i) Initial numbers of the membrane molecules were set proportional to the surface area, thus they were multiplied by the surface-to-volume ratio (1/4 and 1/25 for the volume of 0.8 µm$^3$ and 12.5 µm$^3$, respectively). (ii) Membrane permeability coefficients were also set proportional to the surface area, and thus they were multiplied by the surface-to-volume ratio. (iii) Originally, in the tau-leaping method, constants for the propensity functions of bimolecular reactions are proportional to the inverse of the system volume. Thus, in this model, to set them proportional to the inverse of the surface area, constants for the propensity functions of the bimolecular reactions occurring on the membrane were divided by the surface-to-volume ratio. Likewise, the constants for the propensity functions of the trimolecular reactions were divided by the square of the surface-to-volume ratio. Surface molecules in this model are metabotropic glutamate receptor (mGluR), IP$_3$ receptor (IP$_3$R), plasma membrane Ca$^{2+}$-ATPase (PMCA), sacro- and endoplasmic reticulum Ca$^{2+}$-ATPase (SERCA), Na$^+$/Ca$^{2+}$ exchangers (NCX), and their complexes with other moleucles. Reactions occurring on the membrane were all reactions which involve the surface molecules. As a result, if the numbers of the membrane molecules are set proportional to the surface area, large Ca$^{2+}$ increase did not occur at any PF-CF intervals. **A, B,** Time courses of the concentrations of Ca$^{2+}$ in the volume of 0.8 µm$^3$ (**A**) and 12.5 µm$^3$ (**B**) in response to PF and CF inputs with ˮt = 160 msec. Results of 20 trials of simulation with the numbers of membrane proteins set proportional to the surface area (red lines) and to the cytosolic volume (black lines) were shown. **C, D,** Distribution of Ca$^{2+}$ response in 0.8 µm$^3$ (**C**) and 12.5 µm$^3$ (**D**) under the condition that the numbers of membrane proteins were set proportional to the surface area. The numbers of membrane proteins were set $4/8 = 0.5$ times in $0.8\ \mu m^3$ and $25/125 = 0.2$ times in $12.5\ \mu m^3$ as large as those in (**E**) and (**F**). **E, F,** Distribution of $Ca^{2+}$ response in $0.8\ \mu m^3$ (**E**) and $12.5\ \mu m^3$ (**F**) under the condition that the numbers of membrane proteins were set proportional to the cytosolic volume.

**Figure. S8 | Input timing informations calculated on the assumption that the distribution of PF-CF interval follows Gaussian distribution.** Here we calculated the input timing informations on the assumption that the distribution of PF-CF interval follows Gaussian distribution, and compared them with the input timing information calculated on the assumption that the distribution of PF-CF interval follows uniform distribution. Their similarity demonstrates that the results were qualitatively not affected by the distribution of PF-CF interval. PF-CF intervals were assumed to follow Gaussian distribution with mean of 0 ms and half-width of 300ms (dashed lines with triangles), Gaussian distribution with mean of 200 ms and half-width of 300ms (dotted lines with squares), and uniform distribution (solid lines with circles). **A,** Volume-dependency of the input timing information coded by the total distribution of the $Ca^{2+}$ response. **B,** Volume-dependency of the input timing information per volume. **C,** Relative contribution of the probability component to the input timing information. **D,** Input timing information per volume, coded by the $Ca^{2+}$ response, in a spine (red) and in a cell (black). **E,** Input timing information per volume in a spine (red) and in a cell (black). **Fig. S8A**, **B**, and **C** correspond to **Fig. 3D**, **E**, and **F**, respectively. **Fig. S8D** and **E** correspond to **Fig. 4B** and **D**, respectively.

**Supplemental table 1 | Molecules and their initial number.** This model consists of 56 molecules. Each of them is assumed to exist in one of the following compartment: cytosol, postsynaptic density (PSD), endoplasmic reticulum (ER), or extracellular space, with volumes of $0.1\mu m^3$, $0.02\ \mu m^3$, $0.002\mu m^3$, and $10\ \mu m^3$ in a spine respectively. The initial number of the molecules and the volumes of the compartments are the same as the previous deterministic model.

**Supplemental table 2 | Reactions and their parameters.** This model consists of 43 reactions, one decay (decay of glutamate), one channel permeation ($Ca^{2+}$ permeation through $IP^3R$), two diffusions ($Ca^{2+}$ and $IP_3$ diffusions between cytosol and PSD), and two membrane permeations ($Ca^{2+}$ permeation through the membrane of the cell and that through the membrane of the ER). In the stochastic model, decay of glutamate, two diffusions, and two membrane permeations were implemented as unimolecular reactions. $Ca^{2+}$ permeation through $IP_3R$ was implemented as bimolecular reaction between $Ca^{2+}$ and $IP_3R$. All parameters are the same as the previous deterministic model.

**A**

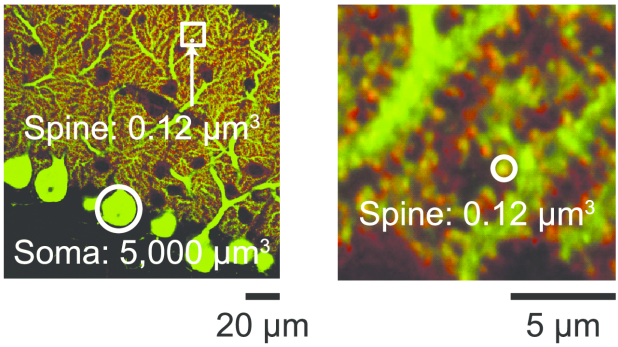

**C**

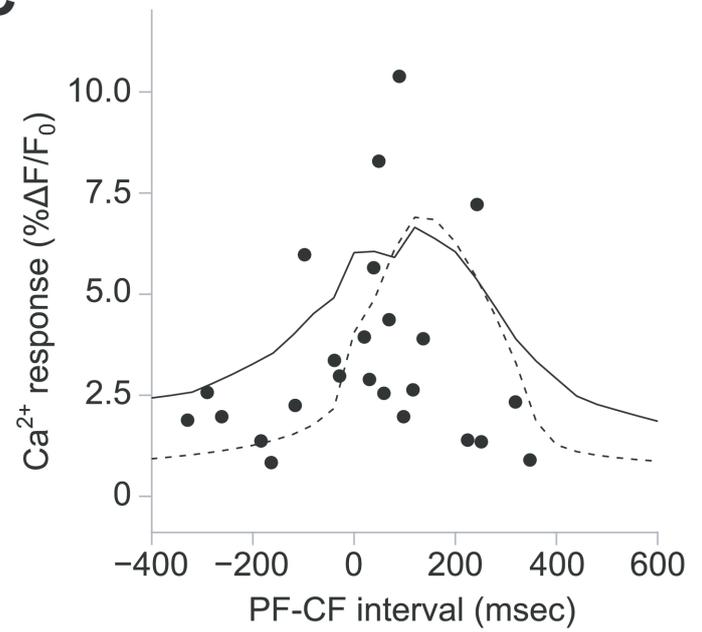

**B**

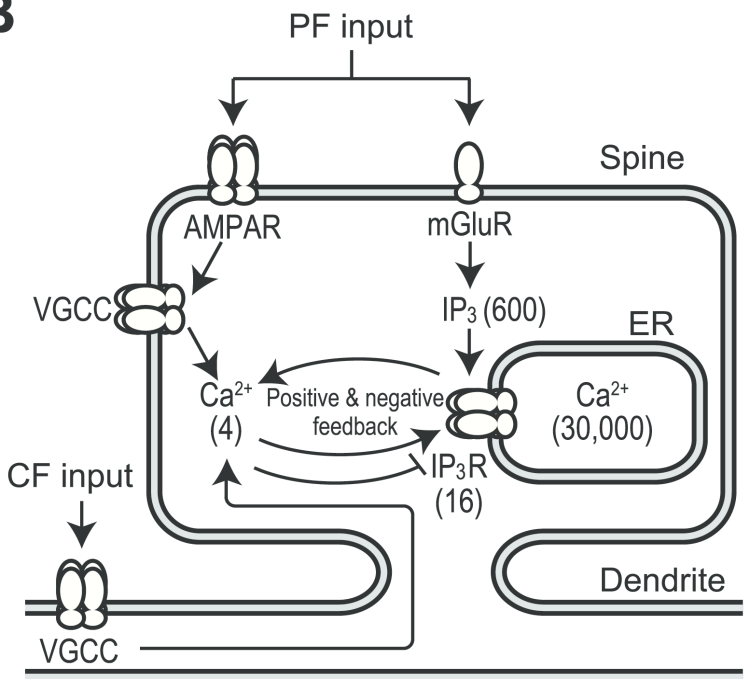

FIgure 1

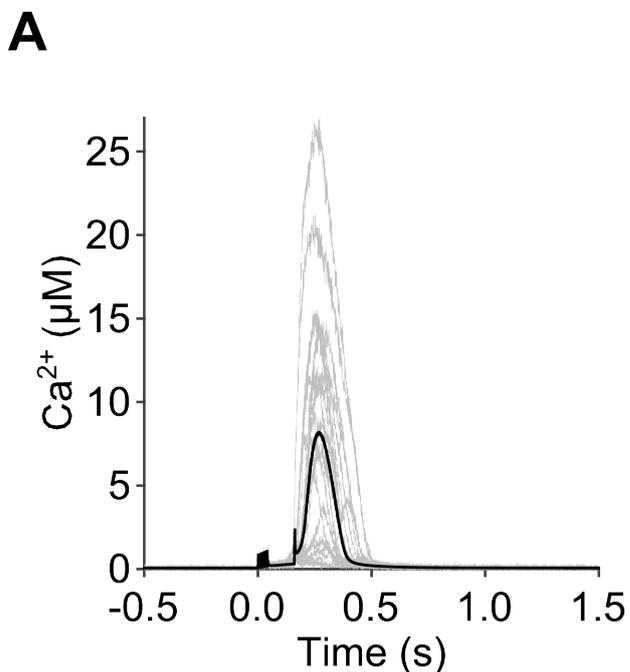

**A**

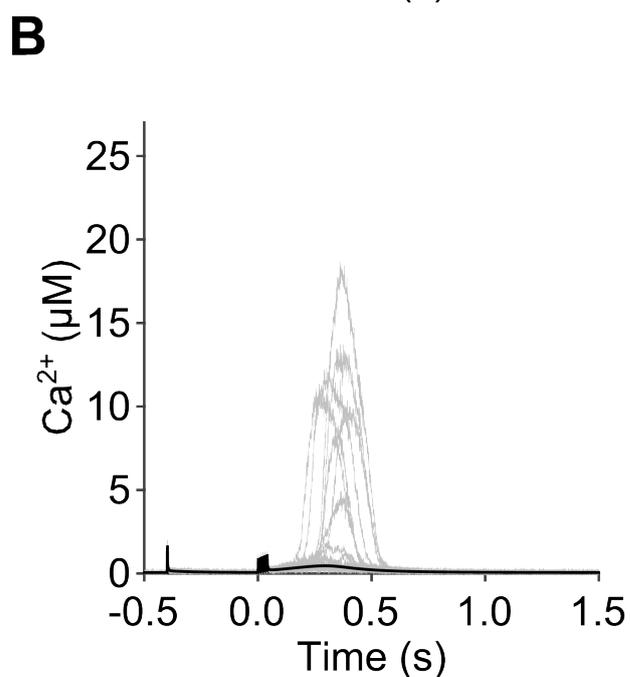

**B**

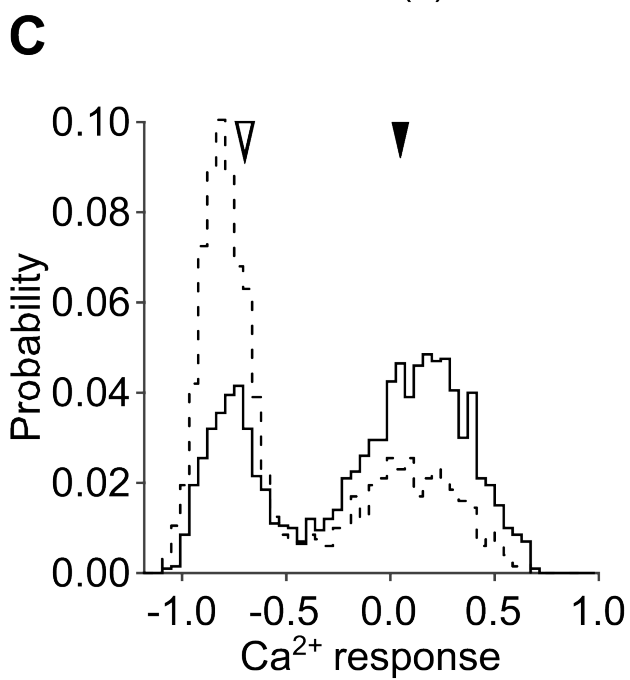

**C**

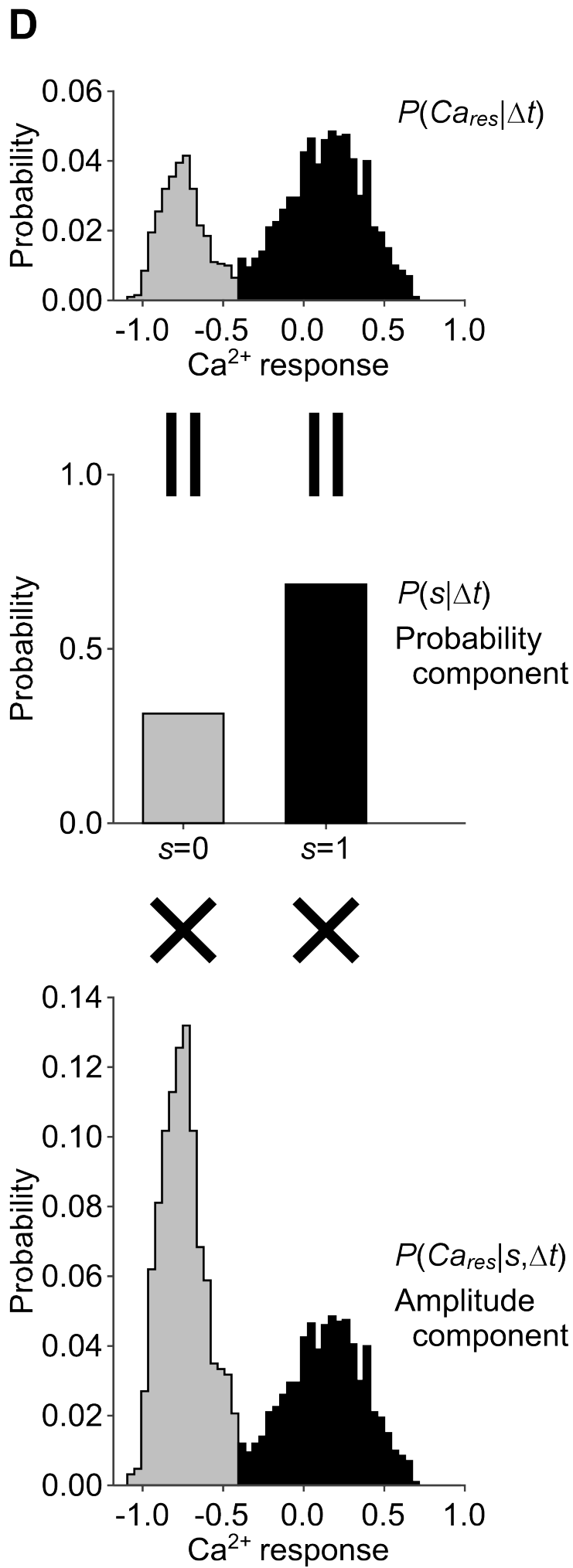

**D**

$P(Ca_{res}|\Delta t)$

$P(s|\Delta t)$
Probability
component

$P(Ca_{res}|s,\Delta t)$
Amplitude
component

Figure 2

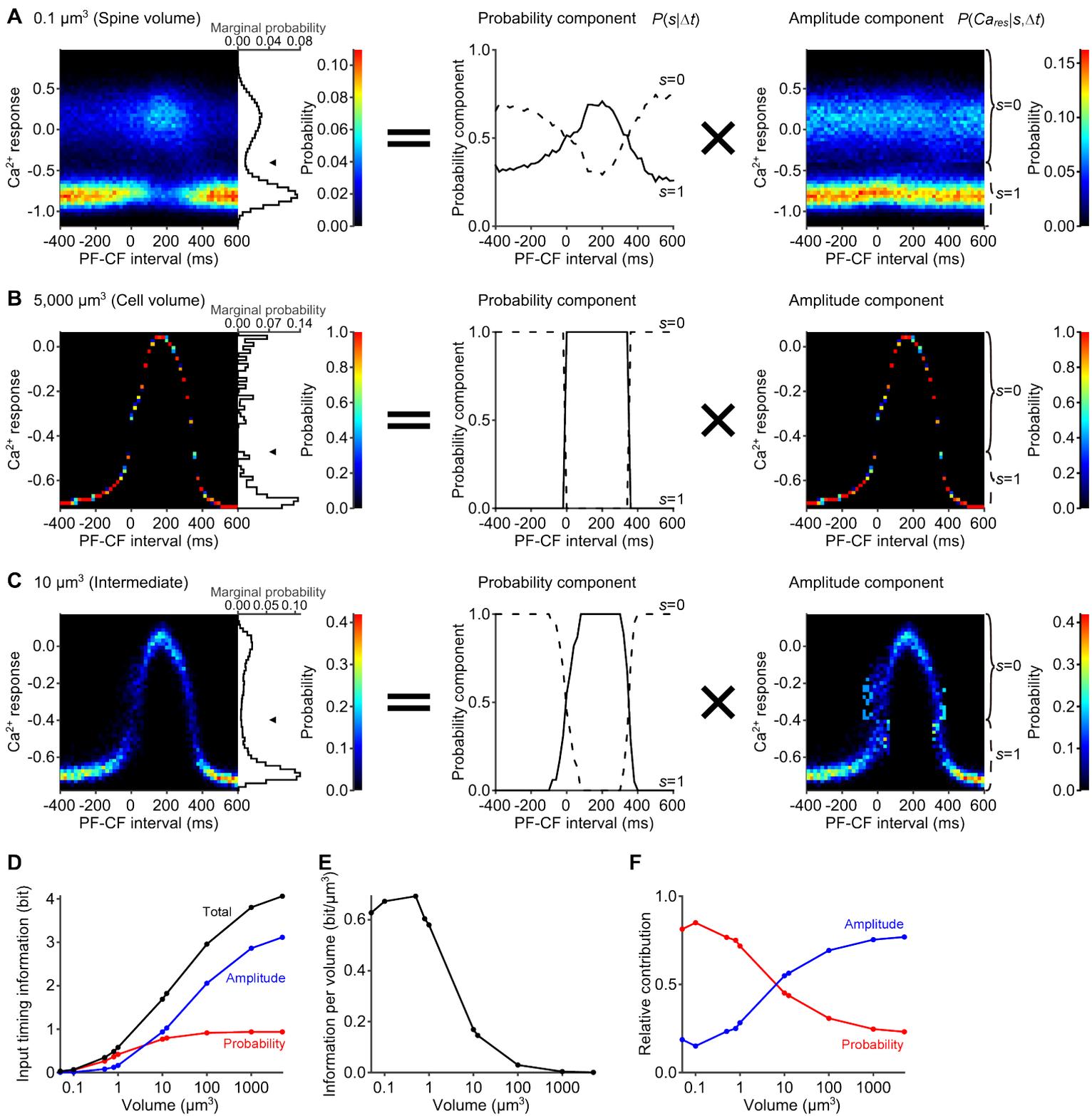

FIgure 3

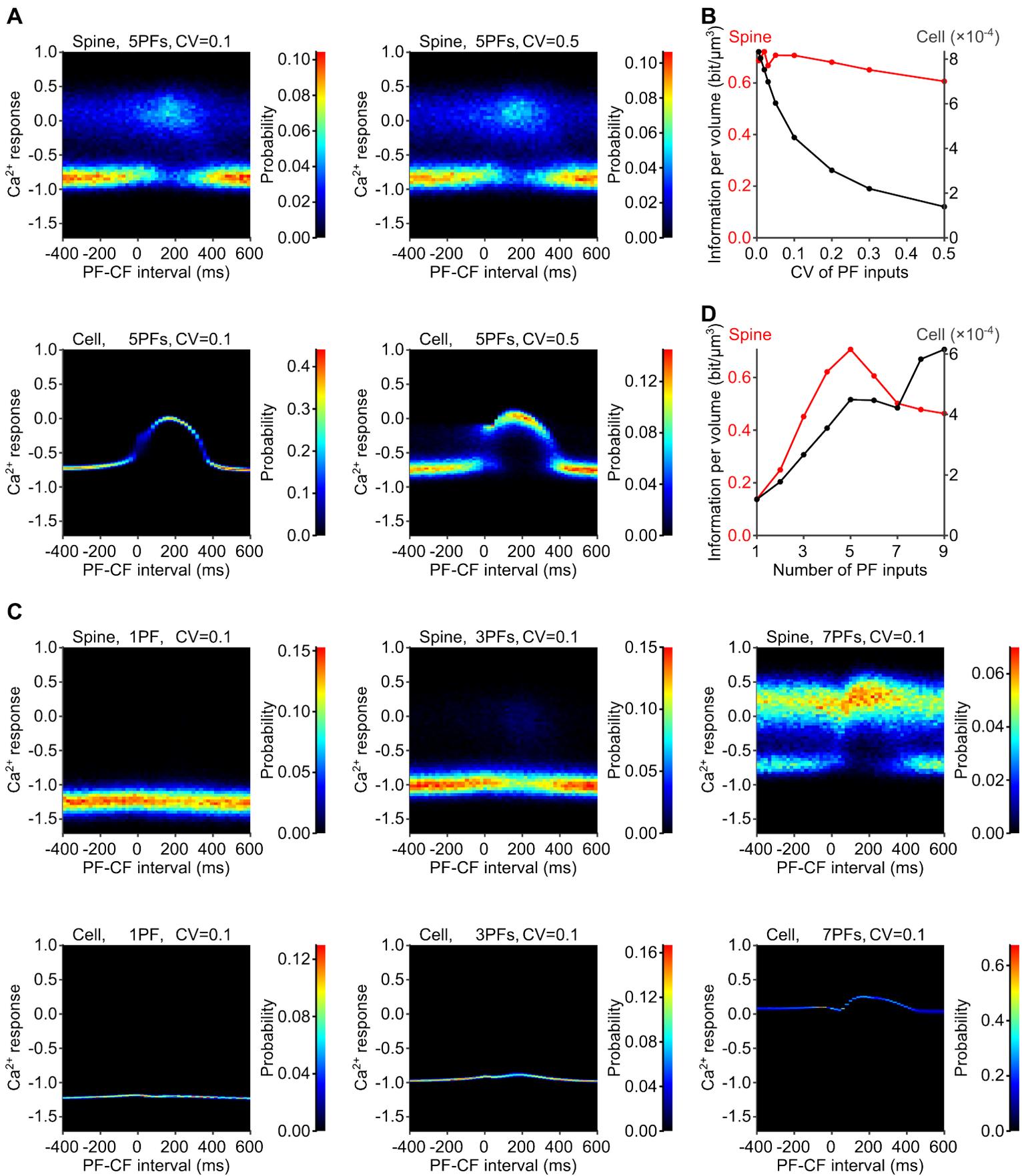

FIgure 4

**A** 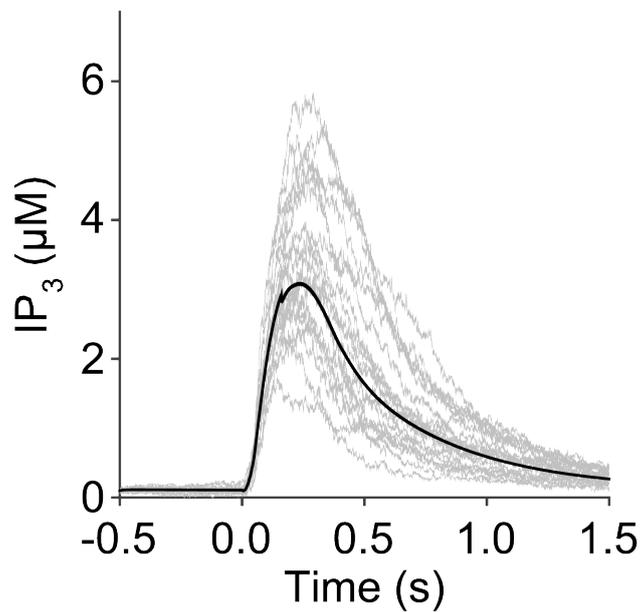

**B** 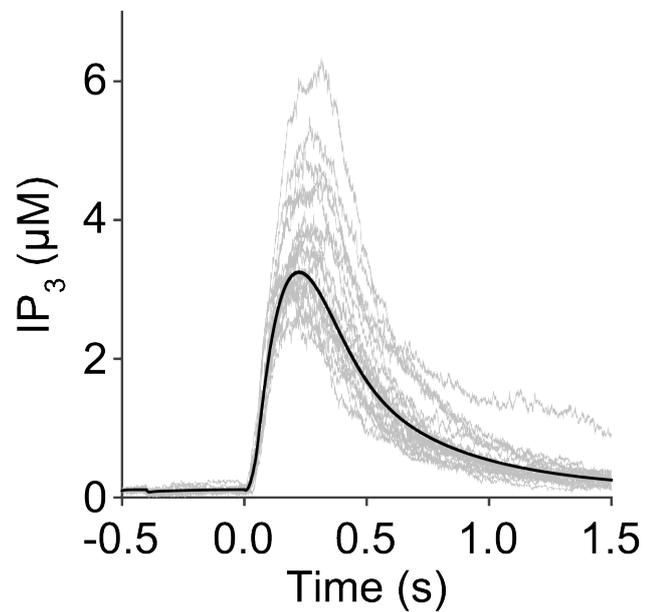

**C** 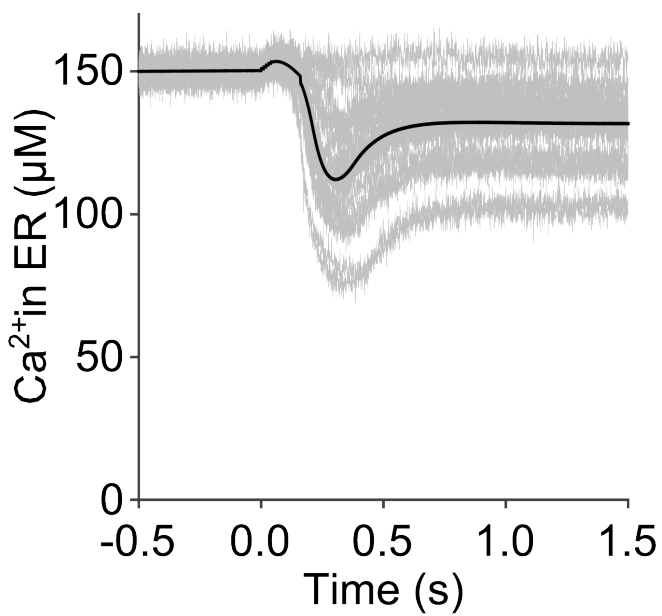

**D** 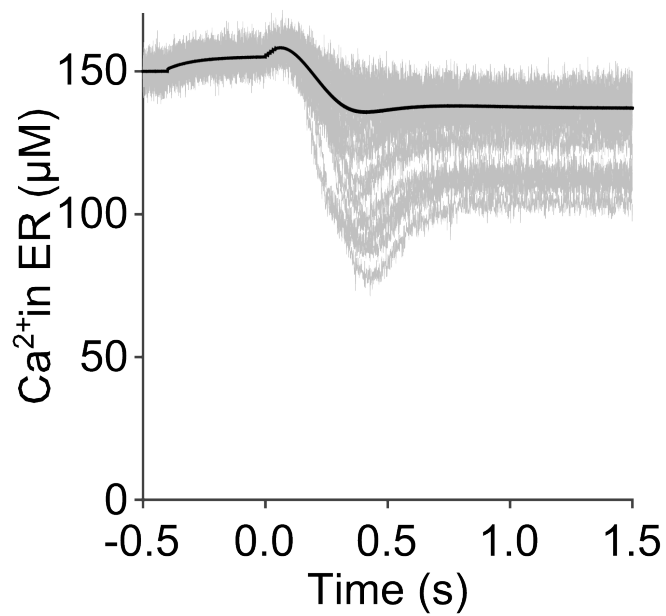

FIgure S1

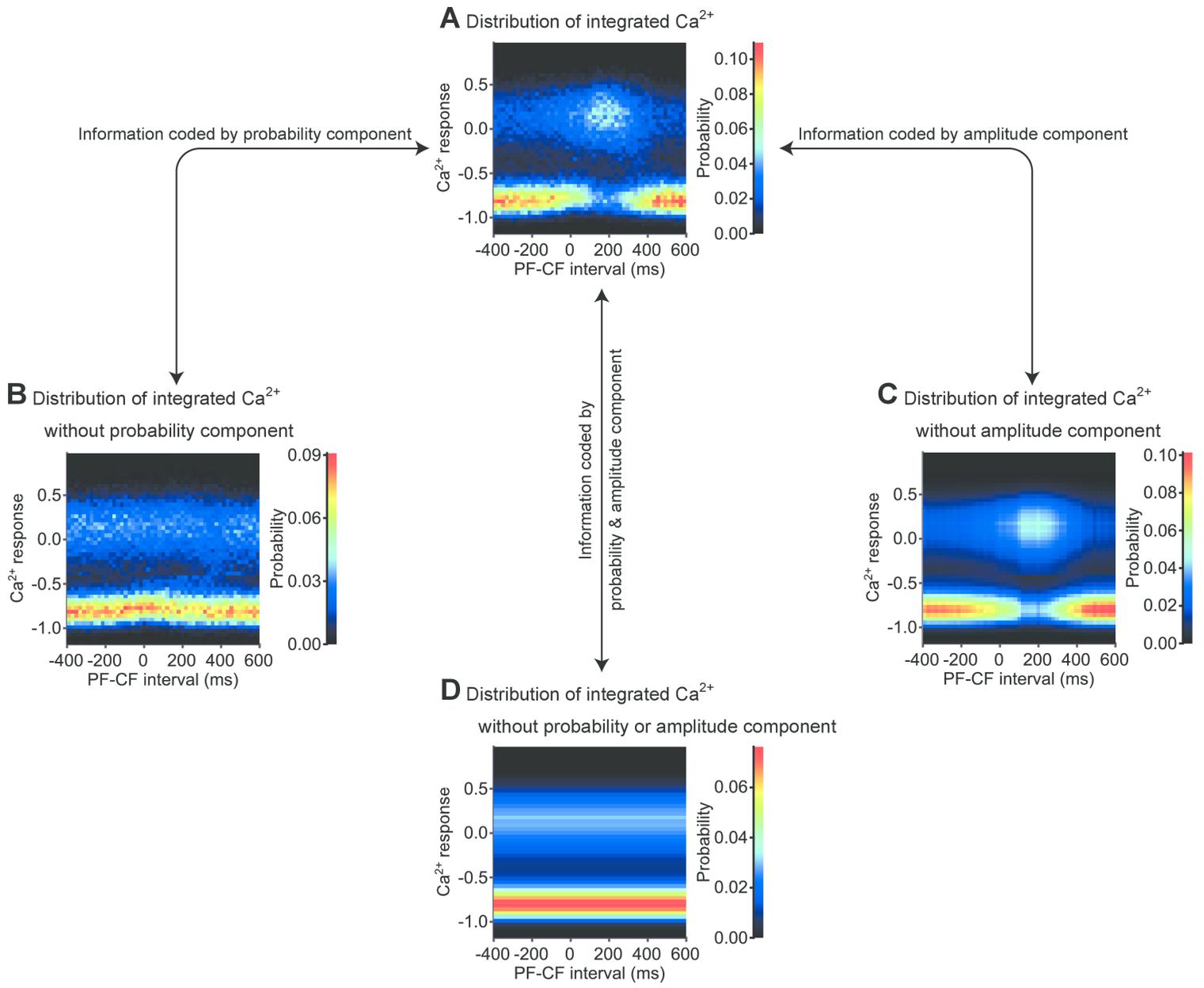

**A** Distribution of integrated Ca²⁺

Information coded by probability component

Information coded by amplitude component

**B** Distribution of integrated Ca²⁺ without probability component

**C** Distribution of integrated Ca²⁺ without amplitude component

Information coded by probability & amplitude component

**D** Distribution of integrated Ca²⁺ without probability or amplitude component

FIgure S2

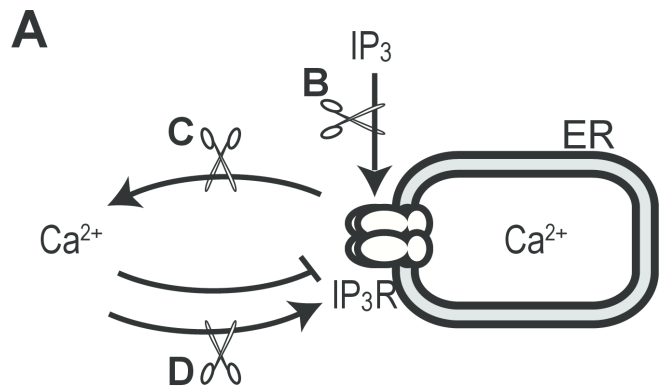

A

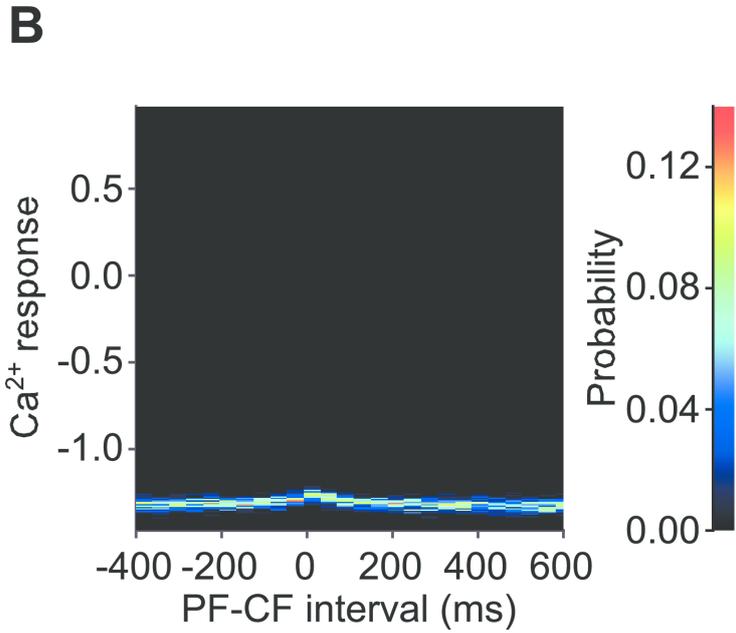

B

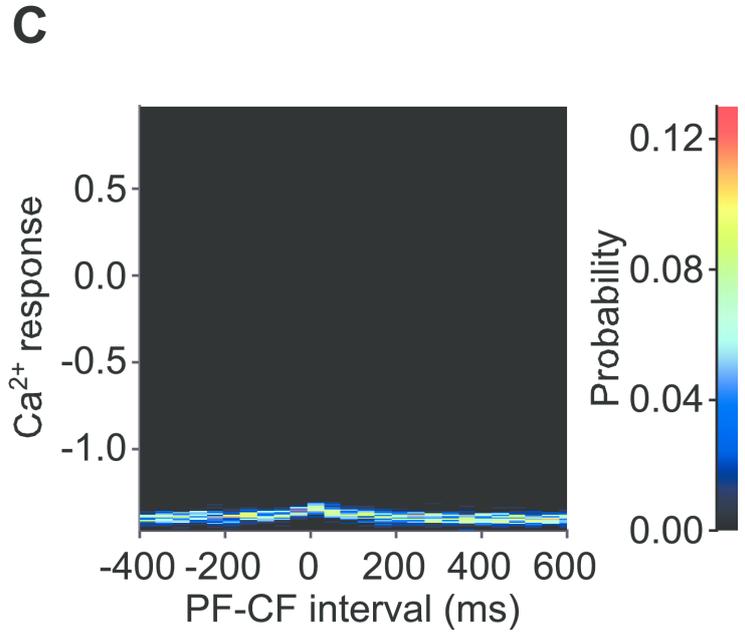

C

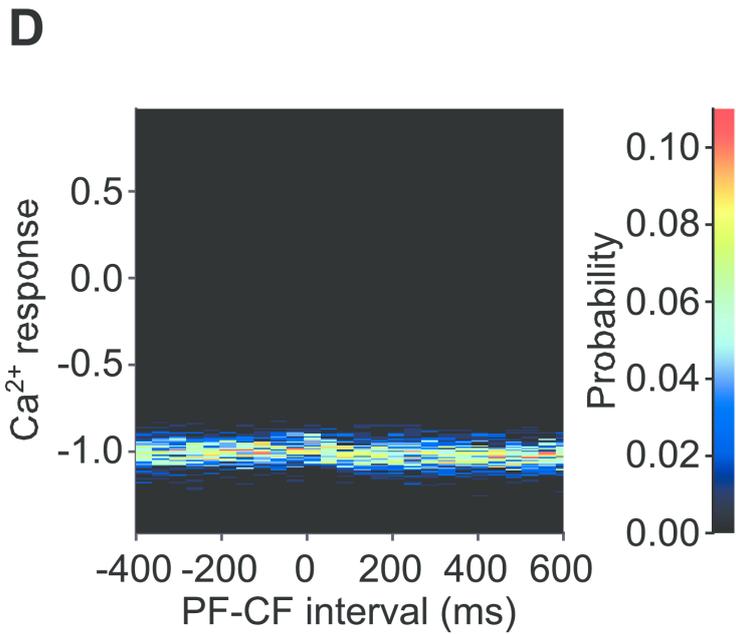

D

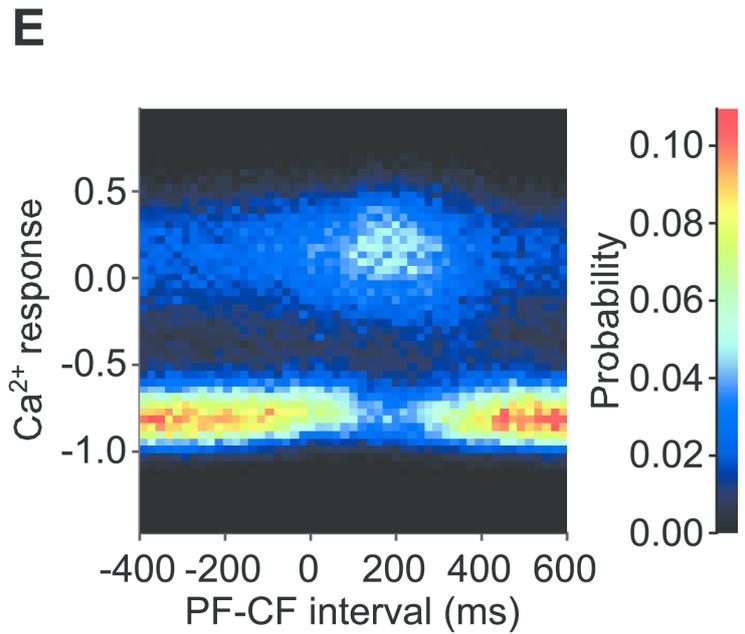

E

Figure S3

**A**

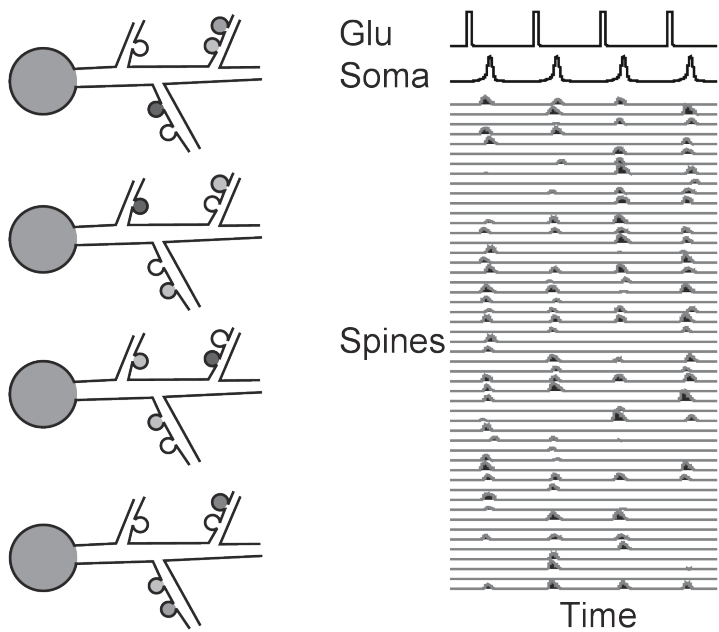

**B**

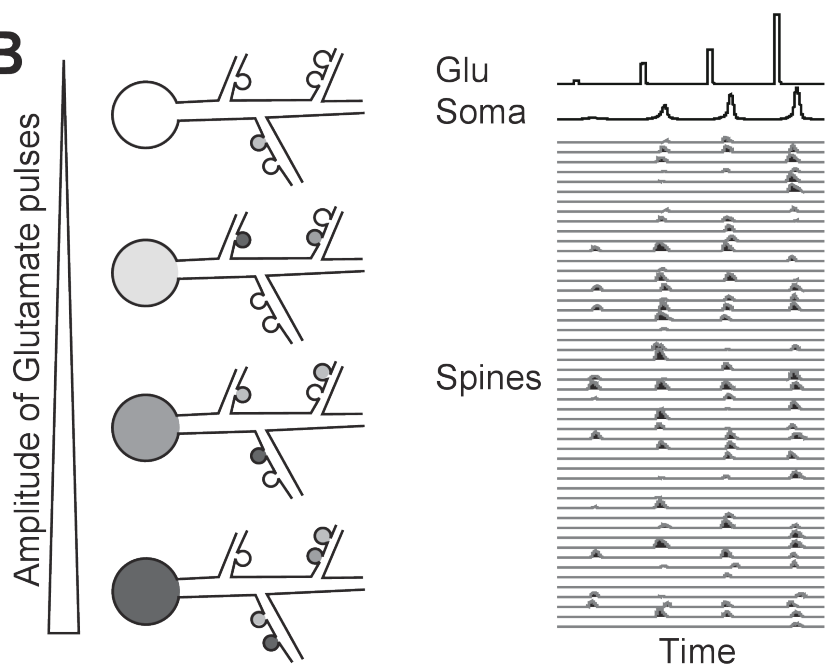

Amplitude of Glutamate pulses

**C**

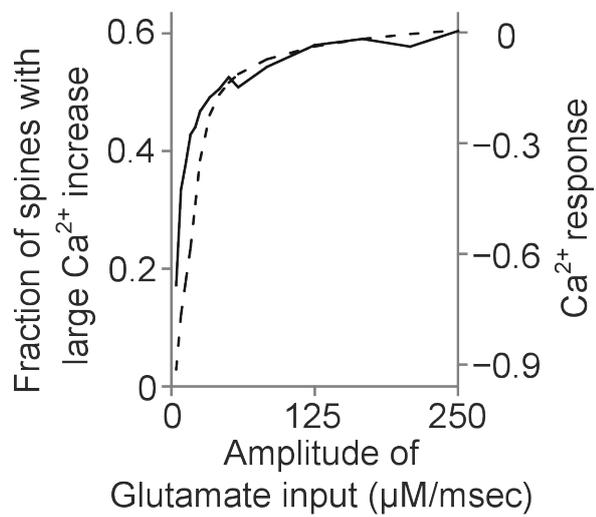

Figure S4

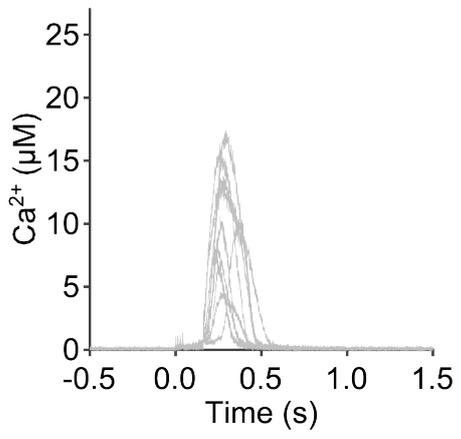

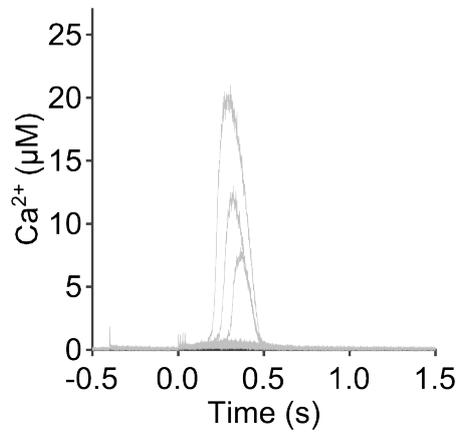

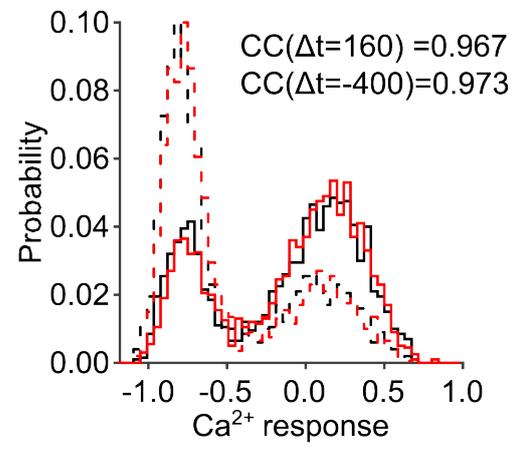

CC(Δt=160) =0.967
CC(Δt=-400)=0.973

FIgure S5

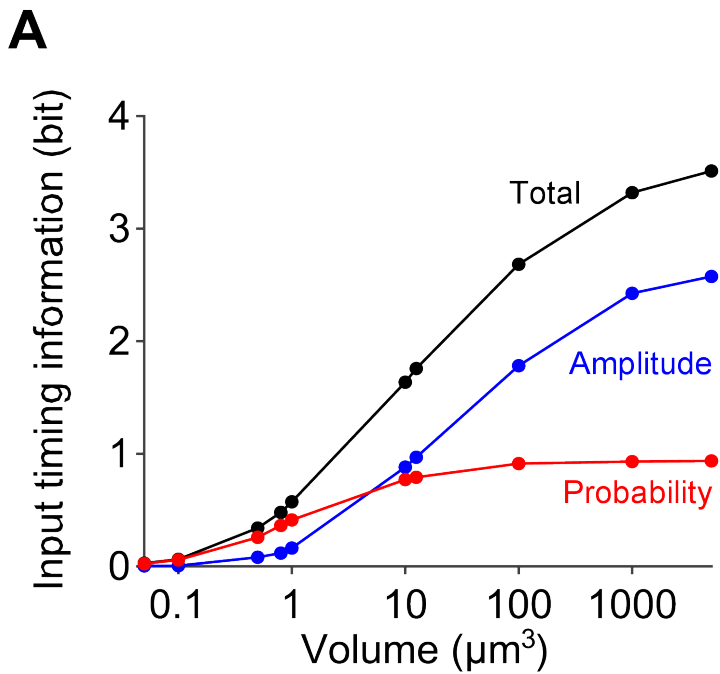

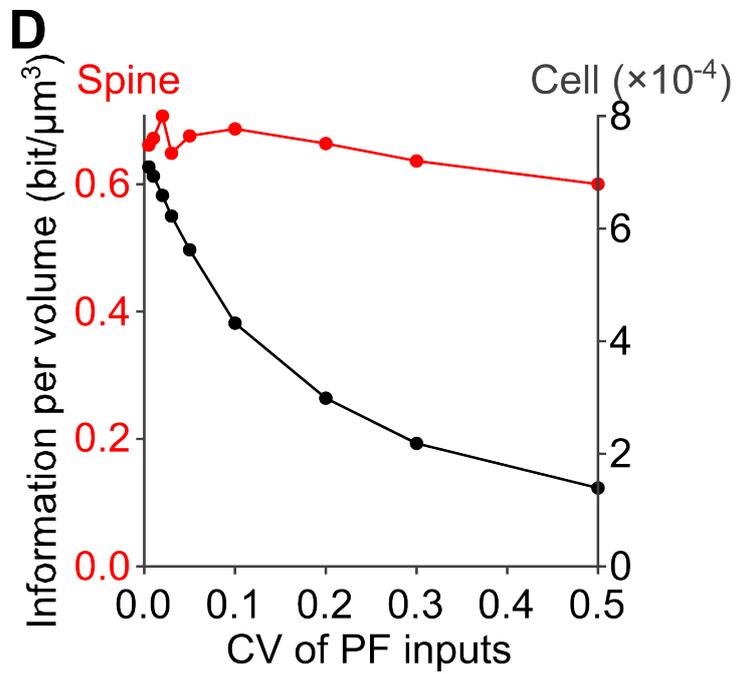

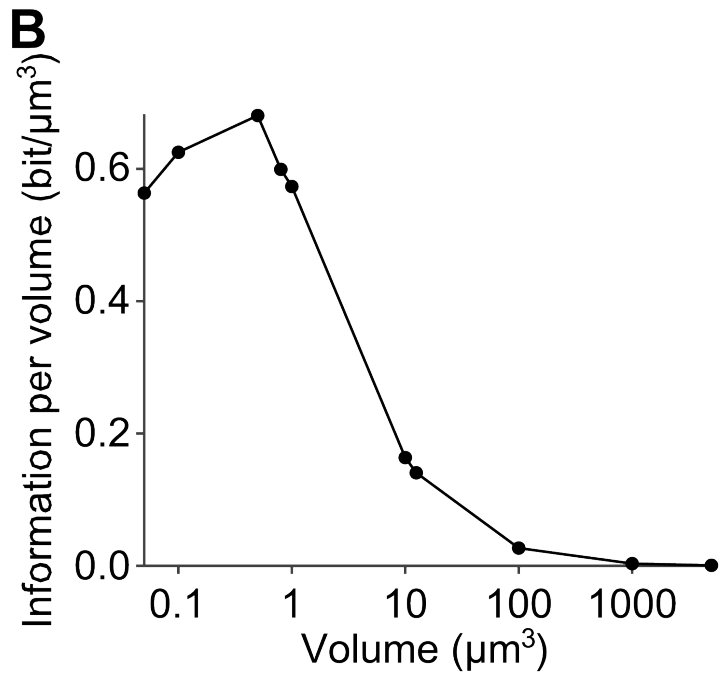

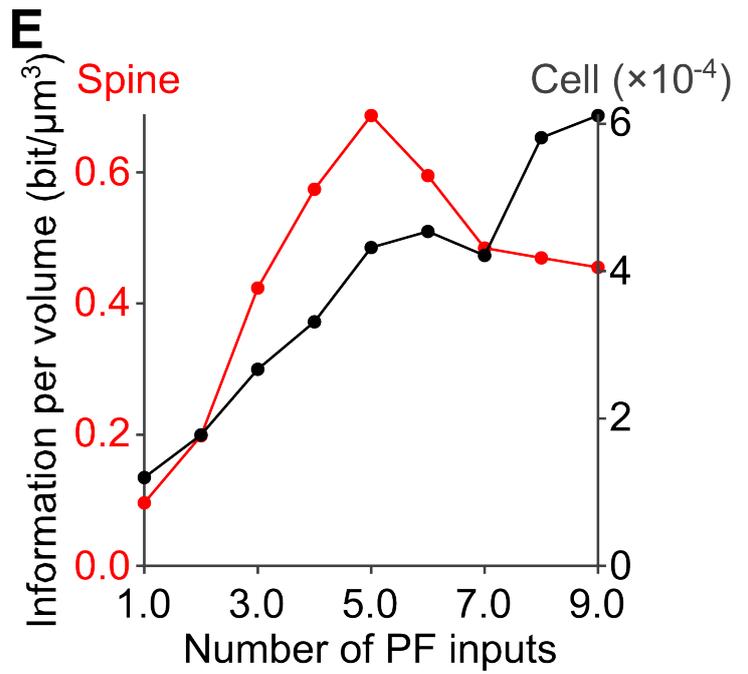

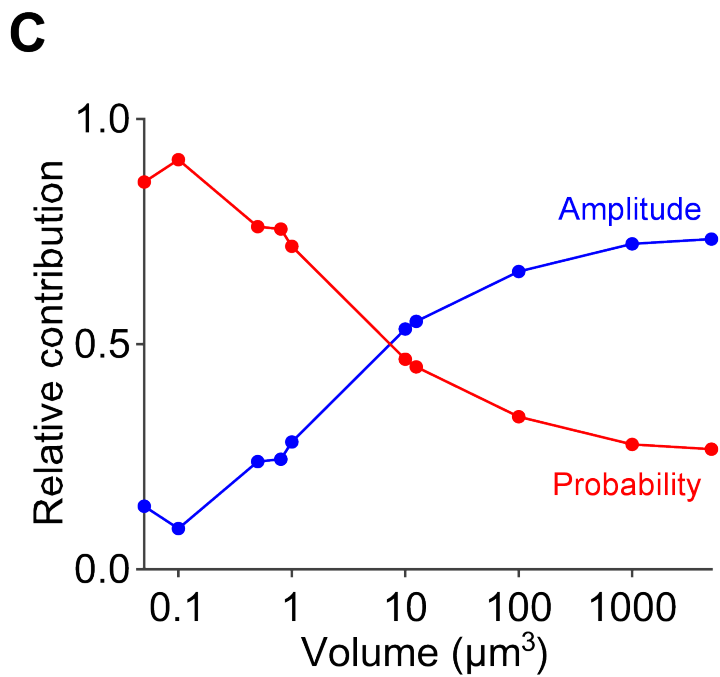

FIgure S6

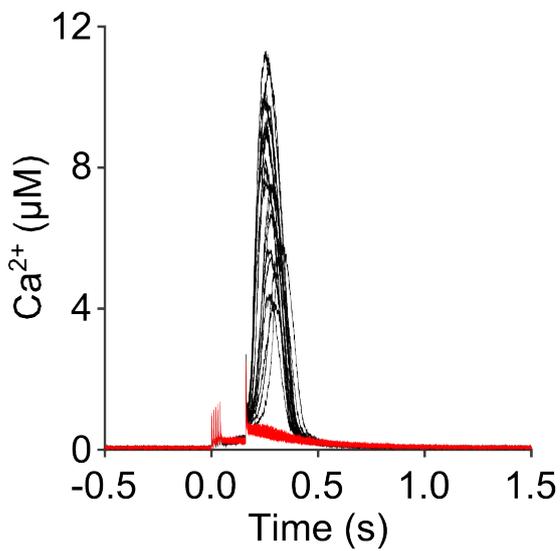

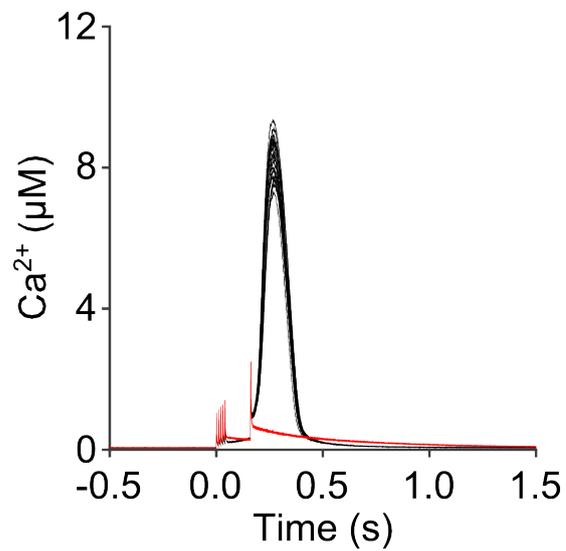

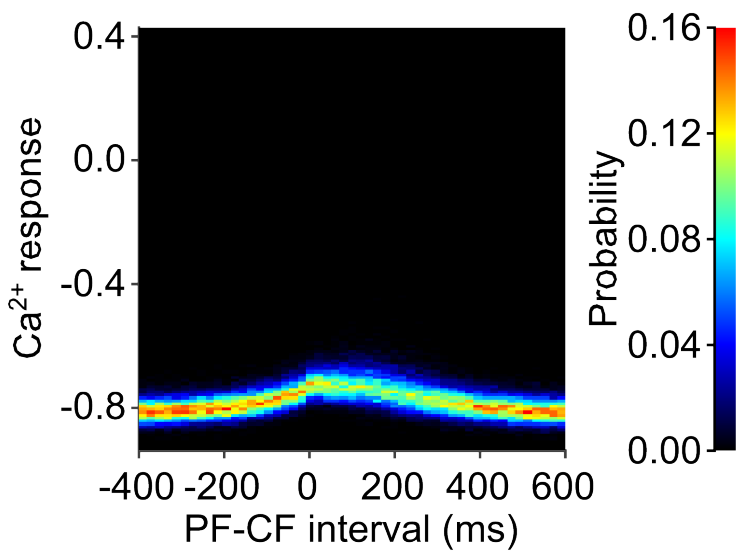

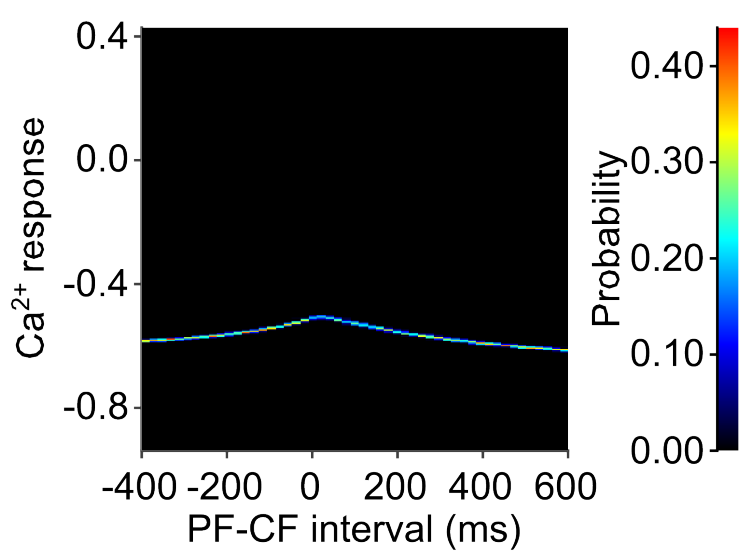

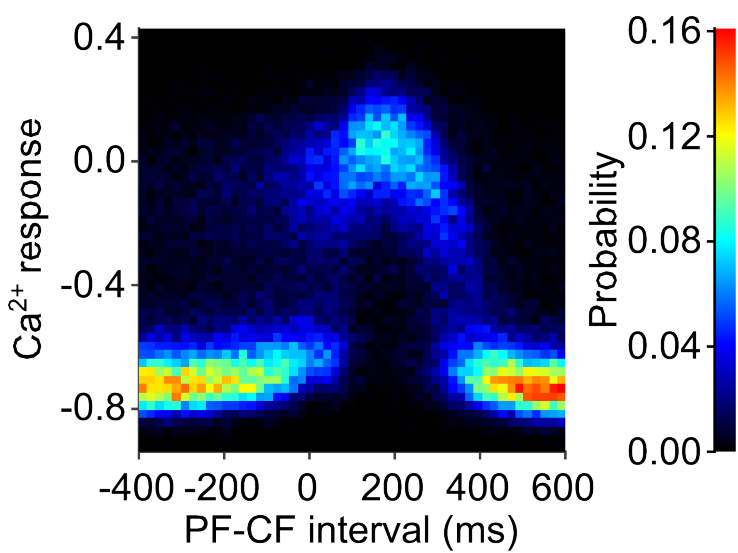

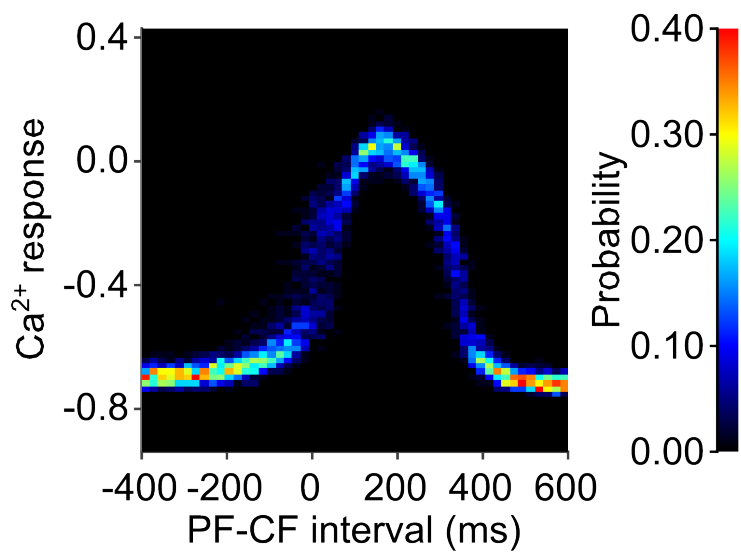

Figure S7

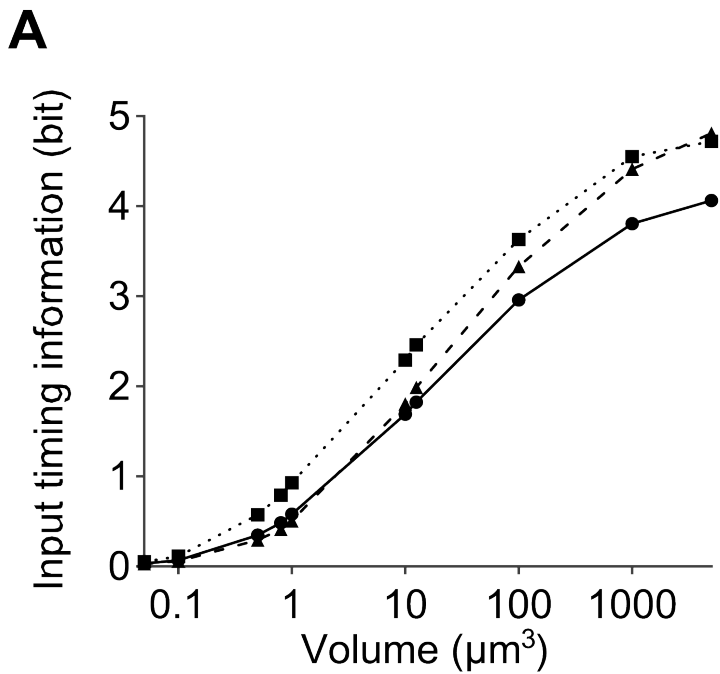

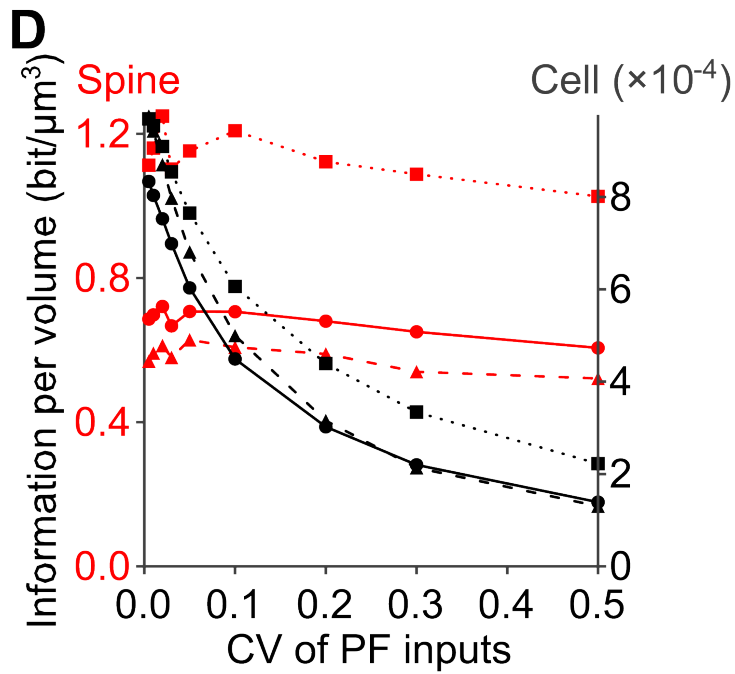

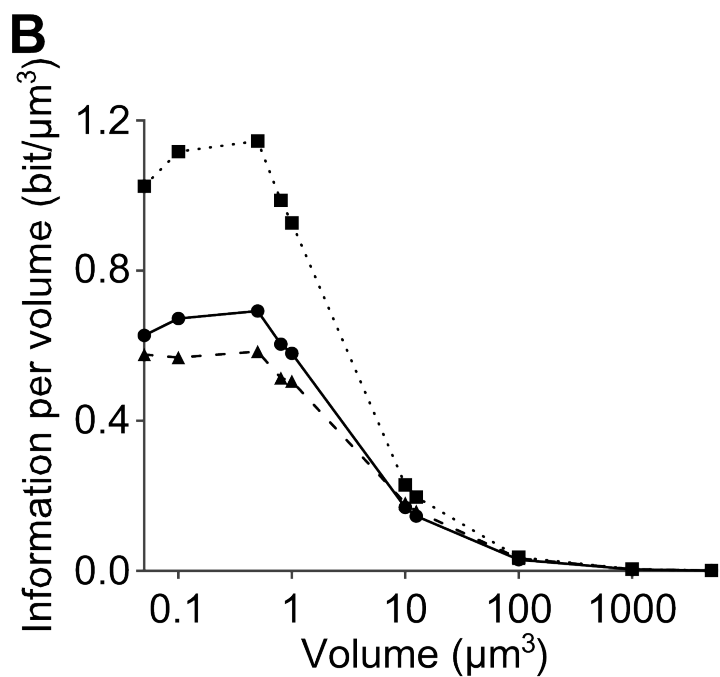

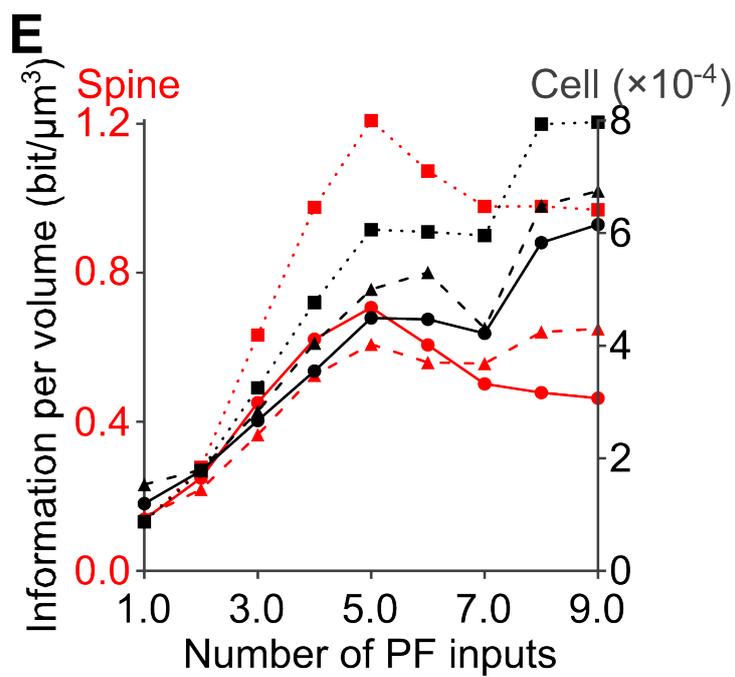

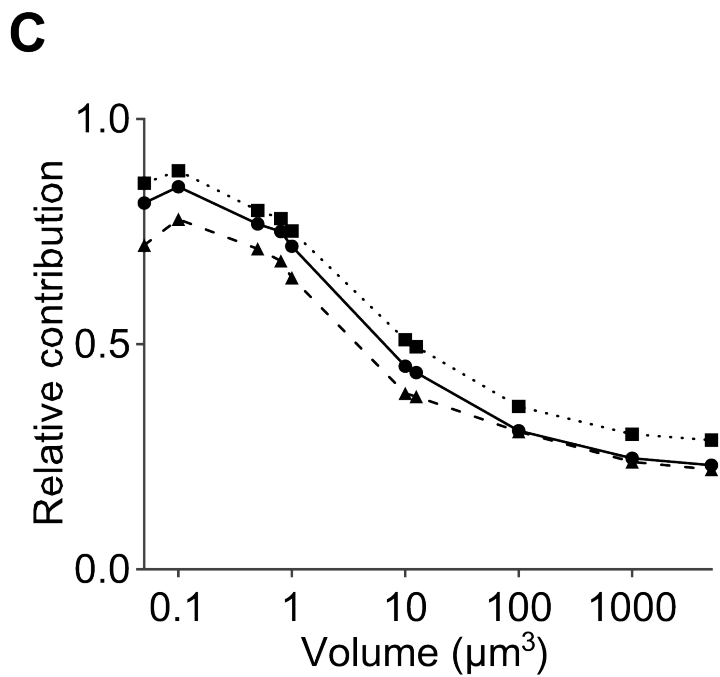

- - - ▲ - - - Gaussian (mean=0 ms, half-width=300 ms)
········■········ Gaussian (mean=200 ms, half-width=300 ms)
———●——— Uniform

FIgure S8